\documentclass[10pt,journal]{IEEEtran}
\IEEEoverridecommandlockouts

\usepackage{graphicx}
\usepackage{subfigure}
\usepackage{amsmath,amsthm,amsfonts,amssymb}
\usepackage{cite}
\usepackage{bm}
\usepackage{bbm}
\usepackage{url}
\usepackage{array}
\usepackage{color}
\usepackage{multirow}
\usepackage{booktabs}
\usepackage[table,xcdraw]{xcolor}
\usepackage{enumitem}

\theoremstyle{plain}
\newtheorem{thm}{Theorem}
\newtheorem{lem}[thm]{Lemma}

\newtheorem{cor}[thm]{Corollary}

\newtheorem{rem}{Remark}
\newtheorem{sty1}{Theorem}
\newtheorem{defi}[sty1]{Definition}

\newenvironment{NewProof}{{\noindent\it Proof.}}{\hfill $\blacksquare$\par}

\usepackage[english]{babel}
\usepackage{algorithm}
\usepackage[noend]{algpseudocode}

\allowdisplaybreaks[4]

\begin{document}
\title{Bayesian Over-the-Air Computation}

\author{
Yulin~Shao,~\IEEEmembership{Member,~IEEE},
Deniz~G\"und\"uz,~\IEEEmembership{Fellow,~IEEE},
Soung Chang Liew,~\IEEEmembership{Fellow,~IEEE}
\thanks{Y. Shao and D. G\"und\"uz are with the Department of Electrical and Electronic Engineering, Imperial College London, London SW7 2AZ, U.K. (e-mail: d.gunduz@imperial.ac.uk). S. C. Liew is with the Department of Information Engineering, The Chinese University of Hong Kong, Shatin, New Territories, Hong Kong (e-mail: soung@ie.cuhk.edu.hk).}
}

\maketitle
\begin{abstract}
As an important piece of the multi-tier computing architecture for future wireless networks, over-the-air computation (OAC) enables efficient function computation in multiple-access edge computing, where a fusion center aims to compute a function of the data distributed at edge devices.
Existing OAC relies exclusively on the maximum likelihood (ML) estimation at the fusion center to recover the arithmetic sum of the transmitted signals from different devices.
ML estimation, however, is much susceptible to noise. In particular, in the misaligned OAC where there are channel misalignments among received signals, ML estimation suffers from severe error propagation and noise enhancement.
To address these challenges, this paper puts forth a Bayesian approach by letting each edge device transmit two pieces of statistical information to the fusion center such that Bayesian estimators can be devised to tackle the misalignments.
Numerical and simulation results verify that,
1) For the aligned and synchronous OAC, our linear minimum mean squared error (LMMSE) estimator significantly outperforms the ML estimator.
In the low signal-to-noise ratio (SNR) regime, the LMMSE estimator reduces the mean squared error (MSE) by at least $6$ dB;
in the high SNR regime, the LMMSE estimator lowers the error floor of MSE by $86.4\%$;
2) For the asynchronous OAC, our LMMSE and sum-product maximum a posteriori (SP-MAP) estimators are on an equal footing in terms of the MSE performance, and are significantly better than the ML estimator. Moreover, the SP-MAP estimator is computationally efficient, the complexity of which grows linearly with the packet length.
\end{abstract}


\section{Introduction}
\begin{figure}[t]
  \centering
  \includegraphics[width=0.9\columnwidth]{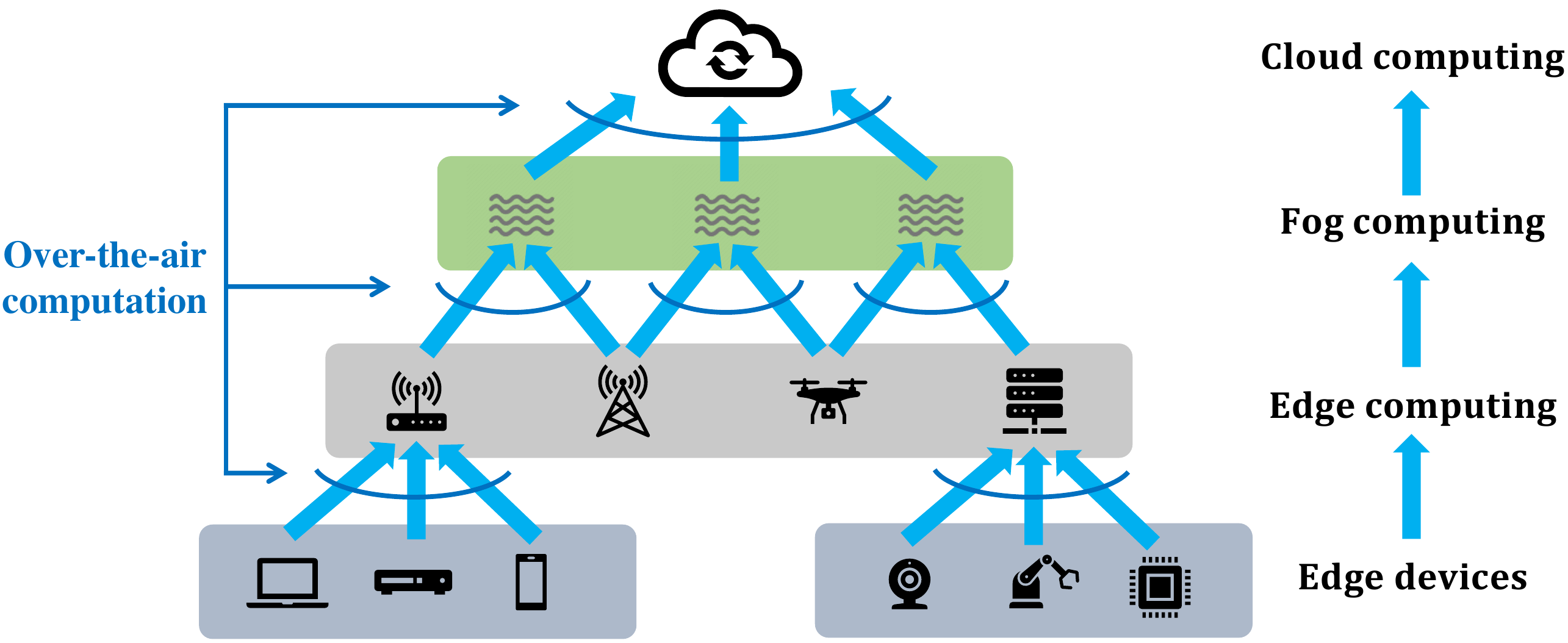}\\
  \caption{Over-the-air computation-enabled multi-tier computing.}
\label{fig:vision}
\end{figure}

Driven by the explosive growth of the number of intelligent devices and their communication and computing demands, future wireless networks are required to exploit the full computing potential of cloud, fog, and edge, forming a new multi-tier computing paradigm \cite{yang2019multi,el2019joint}, as illustrated in Fig.~\ref{fig:vision}. As a promising technique in multi-access edge computing, we envision over-the-air computation (OAC) \cite{Nomographic0,Gunduz1,Zhufirst,ShanghaiTech} becoming an important piece of the new paradigm to provide efficient, scalable, and low-latency function computation services.

In multiple-access edge computing, the fusion center, i.e., the common receiver of the multiple-access channel (MAC), is often interested in some function of the data distributed across the edge devices, rather than their individual values \cite{Gastpar,Nomographic0,Nomographic}.
In distributed sensing networks, for example, the fusion center aims to compute a function of the sensor readings, such as mean humidity or maximum temperature \cite{Robust,zhuNewMag}. In federated learning systems, the fusion center is only interested in the weighted average of the local updates transmitted from the edge devices but not their individual updates \cite{Gunduz1,FedAvg}.

Distributed function computation can be realized in a digital fashion via traditional multiple-access technologies \cite{gupta2015survey,IFDMA} (e.g., TDMA, CDMA, OFDMA). Specifically, the values to be transmitted from the edge devices are first digitized and then transmitted over orthogonal links to the fusion center. The desired function values can be computed after decoding all individual values from the edge devices. However, such a separate communication-and-computation approach is suboptimal in that the individual messages are not the desired targets at the fusion center -- transmitting them causes excessive bandwidth and latency overhead.

OAC is an alternative technique to realize efficient function computation \cite{Gunduz1,FedAvg,ZhuMag,Robust,Nomographic0,zhuNewMag,PNC,Nomographic,zhuPower,techFL,Katabi1,Sery,Gastpar,NoisyNN,ShanghaiTech,Gunduz2,Naifu,amiri2020blind}. Compared with the digital approach, OAC is a joint computation-and-communication scheme exploiting the fact that the MAC inherently generates superposition of signals.
The underpinnings of OAC are pre-processing, channel precoding, and post-processing.
As shown in Fig.~\ref{fig:1}, each device first pre-processes the transmitted symbols by a pre-processing function, and then precodes the pre-processed symbols by the inversion of the uplink channel. The precoded symbols are transmitted to the fusion center in a discrete-time analog fashion \cite{DTAT}. In particular, different devices transmit simultaneously over the same communication link and their signals overlap at the receiver. The fusion center then post-processes the overlapped signal to reconstruct the desired function values.

The pre-processing and post-processing functions are chosen so that the desired function values are directly computed after post-processing. For example, to compute the geometric mean at the fusion center, we can choose the pre-processing function to be a logarithm function and the post-processing function to be an exponential function \cite{Nomographic0}. In general, the functions that are computable via OAC are {\it nomographic functions} \cite{Nomographic} that can be broken into a post-processed summation of multiple pre-processed functions.
On the other hand, the purpose of channel precoding is to compensate for the channel impairments so that the fading MAC degenerates to a Gaussian MAC. As a result, when the transmitted signals arrive at the fusion center simultaneously, the  signals overlapped over the air naturally produce the arithmetic sum of the pre-processed signals.

In practice, however, accurate channel-gain precoding and perfect synchronization among devices are very challenging to achieve \cite{Gunduz1,ZhuMag,V2X}, especially with low-cost Internet-of-Things (IoT) devices.
Therefore, a more interesting and practical setup is the misaligned OAC \cite{techFL}, wherein signals from different devices arrive at the fusion center with either channel-gain mismatches, time asynchronies, or both.

Prior works on the aligned OAC \cite{Gunduz1,Gunduz2,ZhuMag,zhuPower,Sery,zhuNewMag,Nomographic,ShanghaiTech,Simeone,Ozfatura,Katabi1} or the misaligned OAC \cite{techFL} rely exclusively on the maximum likelihood (ML) estimator to recover the arithmetic sum of the transmitted signals from different devices. ML estimation, however, is much susceptible to noise. Our prior work \cite{techFL} showed that, in the misaligned OAC, the arithmetic-sum estimation boils down to multi-user estimation and the estimation space is infinitely large considering the continuous nature of OAC. As a result, ML estimation suffers from severe error propagation and noise enhancement with even mild channel-gain or time misalignment.

This paper puts forth a Bayesian approach for OAC to address the problems faced by ML estimation. Specifically, we let each edge device transmit two pieces of statistical information (i.e., the first and second sample moments) of the distributed data and leverage these statistical characteristics as prior information to construct Bayesian estimators at the receiver to estimate the arithmetic-sum of the transmitted signals. Three OAC systems are considered:
1) The aligned OAC, where the transmitted signals are perfectly aligned at the fusion center with neither channel-gain nor time misalignment;
2) The synchronous OAC, where there is only channel-gain misalignment but no time misalignment. The aligned OAC is a special case of the synchronous OAC when the channel-gain precoding is perfect and there are no channel-gain mismatches in the overlapped signal;
3) The asynchronous OAC, where there are both channel-gain and time misalignments. The synchronous OAC is a special case of the asynchronous OAC when the calibrations of transmission timings at the edge devices are accurate and there are no asynchronies among the overlapped signals.


The main contributions of this paper are as follows:
\begin{enumerate}[leftmargin=0.5cm]
\item We envision an OAC-enabled multi-tier computing paradigm for future wireless networks. We introduce a new ingredient to OAC, i.e., the statistical information about the distributed data at the edge devices, which is conveyed to the fusion center, to design Bayesian estimators that significantly outperform the widely-used ML estimator in OAC.
\item For the aligned OAC and the synchronous OAC, we devise a linear minimum mean square error (LMMSE) estimator using the two pieces of prior information transmitted from the edge devices. The MSE performances of both ML and LMMSE estimators are derived. Numerical results verify that i) for the aligned OAC, the use of prior information brings at least a $6$ dB gain over the ML estimator in the low EsN0 (i.e., the received energy per symbol to noise power spectral density ratio) regime; ii) for the synchronous OAC, the MSE performance of OAC exhibits an error floor due to the misaligned channel coefficients. Compared with the ML estimator, our LMMSE estimator lowers the error floor by a large margin. When there is mild phase misalignment, for example, the error floor is lowered by $86.4\%$ in the high-EsN0 regime.
\item For the asynchronous OAC, we make use of a whitened matched-filtering and sampling scheme to produce oversampled, but independent samples, whereby an ML estimator, an LMMSE estimator, and a sum-product maximum {\it a posteriori} (SP-MAP) estimator are devised, respectively. In particular, our SP-MAP estimator exploits both the prior information transmitted from the edge devices and the sparsity of the sample structure, and is verified to be the most effective estimator in the asynchronous OAC scenario in terms of both the MSE performance and the computational complexity.
Compared with the ML estimator, the SP-MAP estimator addresses the problems of error propagation and noise enhancement and attains significantly lower MSE under various degrees of time and phase misalignments.
While the MSE performance of the SP-MAP estimator is on an equal footing with the LMMSE estimator, it reduces the computational complexity from  $\Omega(L^2\log L)$ to $\Omega(L)$ for a packet of length $L$.
\end{enumerate}

\subsection{Related works}
As an efficient function computation scheme in MACs, OAC finds applications in a variety of scenarios, such as IoT \cite{Nomographic0,Nomographic,Robust,RobustCSI}, federate edge learning \cite{Gunduz1,Gunduz2,Zhufirst,ShanghaiTech}, massive multiple access \cite{zhuMIMO,USTCMIMO,AUS2020}, etc. In this paper, we further integrate OAC into the scope of multi-tier computing.

The early studies of OAC focused on the Gaussian MAC scenario, i.e., the perfectly aligned case with neither channel-gain nor time misalignment.
In \cite{Nomographic0,Nomographic}, the authors demonstrated that the structure of nomographic functions allows the utilization of the interference in Gaussian
MACs for efficient computation at a significantly higher rate than standard schemes.
In \cite{Gunduz1,Zhufirst,Sery}, the authors introduced OAC to federated edge learning for efficient distributed stochastic gradient descent over a Gaussian MAC.

A large part of prior works is devoted to the study of synchronous OAC. That is, symbol-level synchronization among edge devices is assumed and there is only channel-gain misalignment in the received signal \cite{USTCMIMO,zhuMIMO,amiri2020blind,ShanghaiTech,Gunduz2,AUS2020,zhuPower}. In these works, the objective is often to minimize the MSE of the reconstructed arithmetic mean at the receiver.
To this end, prior works often assume multiple antennas at the receiver \cite{ShanghaiTech,amiri2020blind} or multiple input and multiple output (MIMO) \cite{zhuMIMO,USTCMIMO}. The MSE of the estimated arithmetic mean is minimized by jointly optimizing the transmit and receive beamforming, under the assumption that the channels from different edge devices to the receiver are independent and identically distributed (i.i.d.).
The estimators designed in these papers are essentially the ML estimator.
When the number of antennas reduces to $1$ \cite{Gunduz2,AUS2020,zhuPower}, their estimators have the same form as the ML estimator in this paper, which is used as a baseline for our Bayesian estimators. Our results in this paper indicate that incorporating the prior information through Bayesian estimation into MIMO OAC systems can potentially yield further improvements in addition to beamforming.

Two lines of prior works that consider asynchronous OAC are \cite{Robust,RobustCSI} and \cite{techFL}. In \cite{Robust,RobustCSI}, the authors assume that each edge device has a single real value to be transmitted, and modulates this real value by a sequence of random phases.
At the receiver, the signal power of the aligned part is collected by matched filtering in a coded division multiple access (CDMA) fashion.
In so doing, the system is shown to be robust to tiny symbol misalignment -- only coarse synchronization is required to ensure a sufficiently large overlap of symbols from different edge devices. This aligned-sample estimator corresponds to the p-ML estimator discussed later in Section \ref{sec:IVA}.
On the other hand, our previous work \cite{techFL} comprehensively studied ML estimation for asynchronous OAC. Results in \cite{techFL} show that, with the ML estimator, asynchronous OAC systems suffer from severe error propagation and noise enhancement when there is inter-symbol and inter-user interference among the received signals. This motivated us to explore the Bayesian approaches for OAC systems.

{\it Notations}: We use boldface lowercase letters to denote column vectors (e.g., $\bm{\theta}$, $\bm{s}$) and boldface uppercase letters to denote matrices (e.g., $\bm{V}$, $\bm{D}$).
For a vector or matrix, $(\cdot)^\top$ denotes the transpose, $(\cdot)^*$ denotes the complex conjugate, $(\cdot)^H$ denotes the conjugate transpose, and $(\cdot)^\dagger$ denotes the Moore-Penrose pseudoinverse.
$\mathbb{R}$ and $\mathbb{C}$ stand for the sets of real and complex values, respectively. $(\cdot)^\mathfrak{r}$ and $(\cdot)^\mathfrak{i}$ stand for the real and imaginary components of a complex symbol or vector, respectively.
The imaginary unit is represented by $j$.
$\mathcal{C}$ and $\mathcal{CN}$ stand for the real and complex Gaussian distributions, respectively.
The cardinality of a set $\mathcal{V}$ is denoted by $|\mathcal{V}|$.
The sign function is denoted by $\text{sgn}(\cdot)$.
The indicator function is denoted by $\mathbbm{1}$.

\section{System Model}\label{sec:II}
\begin{figure}[t]
  \centering
  \includegraphics[width=1\columnwidth]{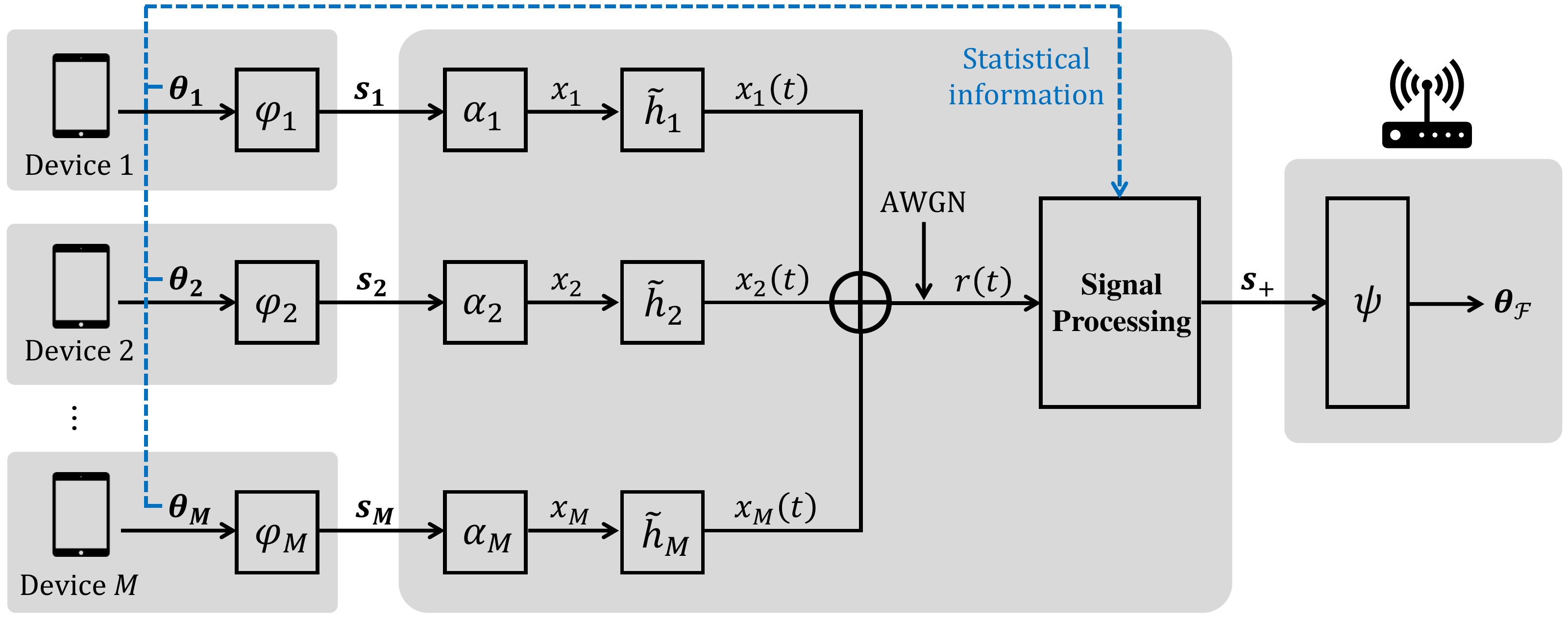}\\
  \caption{Function computation in multi-access edge computing via OAC. With our Bayesian approach, each device transmits two pieces of statistical information to the fusion center.}
\label{fig:1}
\end{figure}
We consider a multi-access edge computing system where $M$ edge devices communicate with a fusion center, as shown in Fig.~\ref{fig:1}. The message of the $m$-th device is a vector of $L$ complex values $\bm{\theta_m}\in\mathbb{C}^L$. The desired message of the fusion center, denoted by $\bm{\theta}_\mathcal{F}\in\mathbb{C}^L$, is a function of $\{\bm{\theta_m}:m=1,2,...,M\}$ and each element of $\bm{\theta}_\mathcal{F}$ is ${\theta}_\mathcal{F}[\ell]=\mathcal{F}(\theta_1[\ell],\theta_2[\ell],...,\theta_M[\ell])$. In particular, $\mathcal{F}$ can be written in a nomographic form as
\begin{eqnarray}\label{eq:II1}
\mathcal{F}(\theta_1[\ell],\theta_2[\ell],...,\theta_M[\ell])=\psi\left(\sum_{m=1}^{M}\varphi_m(\theta_m[\ell])\right),
\end{eqnarray}
where $\{\varphi_m,~m=1,2,...,M\}$ are the pre-processing functions and $\psi$ is a post-processing function. In other words, the nomographic form is a post-processed summation of multiple pre-processed functions. It has been shown in \cite{Buck} that in general any function can be written into a nomographic form.

\subsection{Function computation over the air}
With OAC, function computation over a MAC works in the following manner.
First, each of the $M$ devices pre-processes its message $\bm{\theta_m}$ by a preprocessing function $\varphi_m$ and obtains
\begin{eqnarray}
\bm{s_m}=\varphi_m(\bm{\theta_m}).
\end{eqnarray}
The pre-processed message $\bm{s_m}$ is then precoded by a channel-precoding factor $\alpha_m$ (in the ideal case, $\alpha_m$ corresponds to channel inversion).

After pulse shaping, the time-domain signal to be transmitted by the $m$-th device is given by\footnote{In this paper, we formulate the channel-misaligned OAC considering the time-domain realization of OAC. OAC can also be realized in the frequency domain via OFDM. Interested readers may refer to Section VI of our companion paper \cite{techFL} for a more detailed discussion.}
\begin{eqnarray}\label{eq:II3}
x_m(t)=\alpha_m \sum_{\ell=1}^L s_m[\ell]p(t-\ell T),
\end{eqnarray}
where $p(t)=1/2 \left[\text{sgn}(t+T)-\text{sgn}(t)\right]$ is a rectangular pulse of duration $T$.
Each edge device then carefully calibrates its transmission timing based on its distance from the fusion center and its moving speed, so that the signals from different devices arrive at the fusion center simultaneously.

\begin{figure}[t]
  \centering
  \includegraphics[width=0.9\columnwidth]{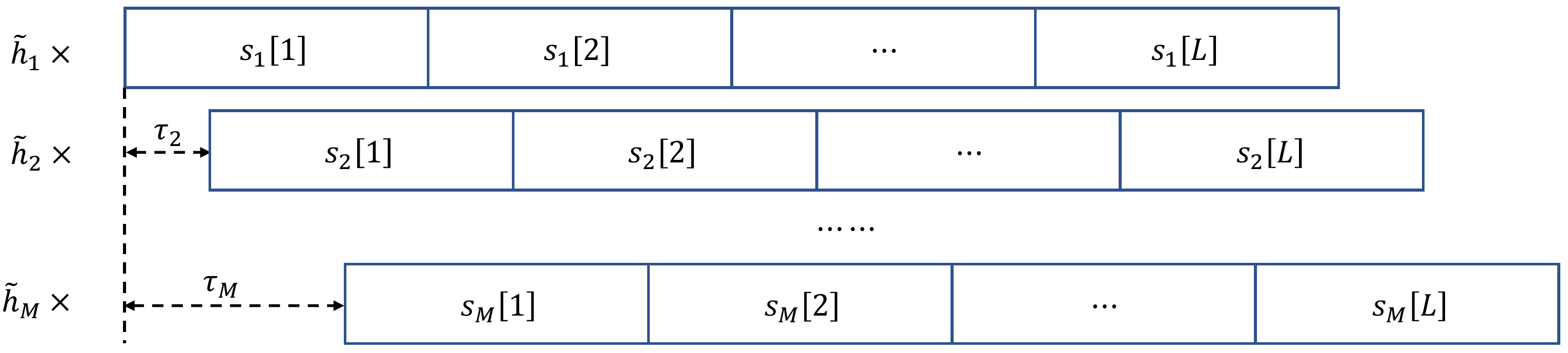}\\
  \caption{The transmitted signals from different devices overlap at the fusion center with channel-gain and time misalignments.}
\label{fig:2}
\end{figure}

In practice, however, both the channel-gain pre-com\-pen\-sa\-tion and the transmission-timing calibration can be imperfect due to the non-ideal hardware and inaccurate channel-ga\-in/delay estimation.
After passing through the fading MAC, the signals $x_m(t)$, $\forall m$,  overlap at the fusion center with relative time offsets. The received signal $r(t)$ can be written as
\begin{eqnarray}\label{eq:II4}
r(t)=\sum_{m=1}^M \widetilde{h}_m x_m(t-\tau_m) +z(t),
\end{eqnarray}
where $\widetilde{h}_m$ is the time-domain complex channel gain. We consider flat and slow fading channels, and hence, $\widetilde{h}_m$ remains constant over one transmission. $z(t)$ is the zero-mean baseband complex additive white Gaussian noise (AWGN), the double-sided power spectral density of which is $N_0$. Without loss of generality, we sort the $M$ devices so that the devices with smaller indexes arrive at the receiver earlier. The delay of the first device is set to $\tau_1=0$, and the relative delay of the $i$-th device with respect to the first device is denoted by $\tau_m$. We assume the time offsets $\tau_m$, $\forall m$, are less than the symbol duration $T$, as shown in Fig.~\ref{fig:2}. In the ideal case where the transmission-timing calibrations are perfect, the relative delays among signals are $\tau_m=0$, $\forall m$.

Substituting \eqref{eq:II3} into \eqref{eq:II4} gives us
\begin{eqnarray}\label{eq:II5}
r(t)\hspace{-0.2cm}&=&\hspace{-0.2cm}\sum_{m=1}^M\widetilde{h}_m\alpha_m\sum_{\ell=1}^L s_m[\ell]p(t-\tau_m-\ell T) + z(t) \nonumber\\
\hspace{-0.2cm}&=&\hspace{-0.2cm} \sum_{\ell=1}^L \sum_{m=1}^M {h}_m s_m[\ell]p(t-\tau_m-\ell T) + z(t),
\end{eqnarray}
where $h_m=\widetilde{h}_m\alpha_m$ is the residual channel gain. To summarize, there can be two kinds of misalignments among the signals transmitted from different devices: channel-gain misalignment $\{h_m:m=1,2,...,M\}$ caused by inaccurate channel-gain compensation at the transmitter, and  time misalignment $\{\tau_m:m=1,2,...,M\}$ caused by imperfect calibration of the transmission timing.

As per \eqref{eq:II1}, the objective of the fusion center is to compute
\begin{eqnarray}
\bm{\theta}_\mathcal{F}=\psi\left(\sum_{m=1}^{M}\varphi_m(\bm{{\theta}_m})\right)\triangleq \psi(\bm{s_+}),
\end{eqnarray}
where each element of sequence $\bm{s_+}$ is defined as
\begin{eqnarray}
s_+[\ell] = \sum_{m=1}^{M} \varphi_m(\theta_m[\ell])=\sum_{m=1}^M s_m[\ell].
\end{eqnarray}

For general pre-processing and post-processing functions, we shall focus exclusively on the estimation of $\bm{s_+}$ from the received signal $r(t)$. In particular, our goal is to minimize the mean squared error (MSE) between the true $\bm{s_+}$ and the estimated $\bm{\widehat{s}_+}$:
\begin{eqnarray}
\text{MSE}(\bm{s_+},\bm{\widehat{s}_+}) = \frac{1}{L}\sum_{\ell=1}^L \left|\widehat{s}_+[\ell]- \sum_{m=1}^M s_m[\ell]\right|^2.
\end{eqnarray}

\begin{rem}
It is worth noting that, for a specific pair of pre-processing and post-pr\-ocessing functions $\varphi_m$ and $\psi$, minimizing the MSE between $\bm{s_+}$ and $\bm{\widehat{s}_+}$ does not necessarily minimize the MSE between $\bm{{\theta}}_\mathcal{F}$ and the estimated $\bm{\widehat{\theta}}_\mathcal{F}$:
\begin{equation}
\text{MSE}(\bm{{\theta}}_\mathcal{F},\bm{\widehat{\theta}}_\mathcal{F}) = \frac{1}{L}\sum_{\ell=1}^L \left|{\widehat{\theta}}_\mathcal{F}[\ell]- \psi\left(\sum_{m=1}^{M}\varphi_m(\theta_m[\ell])\right)\right|^2.
\end{equation}
If the fusion center aims to compute the arithmetic mean of $\{\bm{\theta_m}:m=1,2,..,M\}$, for example, we choose $\varphi_m(x)=x$, $\psi(x)=x/M$, and
\begin{equation}
\text{MSE}(\bm{{\theta}}_\mathcal{F},\bm{\widehat{\theta}}_\mathcal{F}) = \frac{1}{M^2}\text{MSE}(\bm{s_+},\bm{\widehat{s}_+}).
\end{equation}
Therefore, minimizing $\text{MSE}(\bm{{\theta}}_\mathcal{F},\bm{\widehat{\theta}}_\mathcal{F})$ is equivalent to minimizing $\text{MSE}(\bm{s_+},\bm{\widehat{s}_+})$ in this case.

On the other hand, if the fusion center aims to compute the geometric mean of $\{\bm{\theta_m}:m=1,2,..,M\}$, we choose $\varphi_m(x)=\ln x$, $\psi(x)=\exp(x/M)$, and
\begin{equation}
\text{MSE}(\bm{{\theta}}_\mathcal{F},\bm{\widehat{\theta}}_\mathcal{F}) \!=\! \frac{1}{L}\sum_{\ell=1}^L \left|\exp\!\left(\!\frac{\widehat{s}_+[\ell]}{M}\!\right) \!-\!\exp\left(\!\frac{\sum_{m=1}^M \!s_m[\ell]}{M}\right) \right|^2.
\end{equation}
In this case, $\text{MSE}(\bm{{\theta}}_\mathcal{F},\bm{\widehat{\theta}}_\mathcal{F}) \approx\allowbreak \frac{1}{M^2}\text{MSE}(\bm{s_+},\bm{\widehat{s}_+})$ only when both $\frac{1}{M}\allowbreak\widehat{s}_+\allowbreak[\ell]\allowbreak\to 0$ and $\frac{1}{M}\sum_{m=1}^M s_m[\ell]\allowbreak\to 0$.

In this paper, we consider general pre-processing and post-processing functions and focus on the MSE between $\bm{\widehat{s}_+}$ and $\bm{\widehat{s}_+}$.
\end{rem}

\subsection{The Aligned OAC}
Most prior works on OAC considered only the perfectly aligned case, where there is neither channel-gain misalignment nor time misalignment, which we refer to as the {\it aligned OAC}. The received signal in the aligned OAC is given by
\begin{equation}\label{eq:II6}
r(t)=\sum_{\ell=1}^L\sum_{m=1}^M s_m[\ell]p(t-\ell T) +z(t).
\end{equation}
Matched filtering $r(t)$ by the same rectangular pulse $p(t)$ and sampling at $t=iT$, $i=1,2,...,L$, gives us
\begin{equation}\label{eq:II7}
r[i]\!=\!\frac{1}{T}\int_{(i-1)T}^{iT}\!\!r(t) dt = \!\sum_{m=1}^{M}\! s_m[i] \!+\! z[i] = s_+[i] \!+\! z[i],
\end{equation}
where the noise sequence $z[i]$ in the samples is i.i.d., $z[i]\sim\mathcal{CN}(0,\frac{N_0}{T})$.
As can be seen, the target signal $s_+[i]$ appears explicitly on the RHS of \eqref{eq:II7}. A simple estimator can then be used to estimate the sequence $\bm{s}_+$.

\begin{defi}[ML estimation for the aligned OAC]\label{defi:1}
Given a sequence of samples $\bm{r}\in\mathcal{C}^L$ in \eqref{eq:II7}, an ML estimator estimates the target sequence $\bm{s}_+\in\mathcal{C}^L$ symbol-by-symbol by
\begin{eqnarray}\label{eq:II8}
\widehat{s}_+[i]=r[i].
\end{eqnarray}
\end{defi}

Eq. \eqref{eq:II8} is an ML estimator because the sample $r[i]$ in \eqref{eq:II7} is conditional Gaussian -- given $s_+[i]$, the likelihood function
$f(r[i]\left.\right| s_+[i])\sim\mathcal{CN}\left(s_+[i], \frac{N_0}{T}\right)$.
Therefore, the ML estimate of $s_+[i]$ for a given observation $r[i]$ is
$\widehat{s}_+[i]=\arg\max_{s_+[i]}\Pr\left(r[i]\left.\right| s_+[i]\right) = r[i]$.

Accurate precoding and transmission-timing calibration at the transmitters admit a very simple sample structure at the receiver since the target signal $\bm{s}_+$ is explicitly presented in the samples. In this case, the fading MAC degenerates to a Gaussian MAC and the $M$ devices can be abstracted as a single device transmitting the summation of $\bm{s_m}$ directly to the fusion center.
In practice, however, both the channel-gain compensation and the calibration of transmission timing can be inaccurate. With either channel-gain or time misalignment, clean samples as in \eqref{eq:II7} with $\bm{s}_+$ explicitly presented are no longer available. The design of optimal estimators thus becomes more challenging.


\section{Synchronous OAC}\label{sec:III}
This section focuses on the synchronous OAC where there is channel-gain misalignment but no time misalignment in the received signal.
That is, we assume the calibrations of transmission timing at the edge devices are satisfactory and the relative time offsets of different signals at the fusion center are negligible.

After matched filtering and sampling, the samples we obtained can be written as
\begin{eqnarray}\label{eq:III1}
r[i] = \sum_{m=1}^{M}h_m s_m[i] + z[i],
\end{eqnarray}
where $h_m$ is the residual channel gain due to inaccurate chan\-nel-gain precoding. From each sample $r[i]$, our goal is to estimate ${s}_+[i]=\sum_{m=1}^{M}s_m[i]$.

Eq. \eqref{eq:III1} is an underdetermined equation since we have one equation for $M$ unknowns. To estimate ${s}_+[i]$, the only viable estimator in the literature is the ML estimator given in Definition \ref{defi:1} -- the raw sample $r[i]$ is the best prediction about ${s}_+[i]$ \cite{techFL}.
In this paper, however, we will show that the MSE of the estimated $\bm{\widehat{s}_+}$ can be significantly reduced by a Bayesian approach.
Unlike ML estimation where the transmitted symbols $\bm{s}_m$ are treated as constants, we treat $\bm{s}_m$ as random variables (with unknown priors) and exploit the statistical characteristic of $\bm{s}_m$ as a kind of prior information to perform Bayesian estimation at the receiver.
\subsection{ML estimation for the synchronous OAC}
To start with, let us analyze the performance of the ML estimator for the synchronous OAC.

As per Definition \ref{defi:1}, we have $\widehat{s}_+[i] = r[i]$ with the ML estimator.
From \eqref{eq:III1}, the MSE of $\widehat{\bm{s}}_+$ is
\begin{eqnarray*}
\text{MSE}_\text{ML} \hspace{-0.2cm}&=&\hspace{-0.2cm} \frac{1}{L}\sum_{i=1}^{L}\left|\widehat{s}_+[i]-{s}_+[i]\right|^2 \\
\hspace{-0.2cm}&=&\hspace{-0.2cm} \frac{1}{L}\sum_{i=1}^{L}\left|\sum_{m=1}^M (h_m-1)s_m[i]+z[i]\right|^2 \\
\hspace{-0.2cm}&\overset{(a)}{=}&\hspace{-0.2cm} \frac{1}{L}\sum_{i=1}^{L}\left|\sum_{m=1}^M (h_m-1)s_m[i]\right|^2 + \frac{1}{L}\sum_{i=1}^{L}\left|z[i]\right|^2 \\
\hspace{-0.2cm}&=&\hspace{-0.2cm} \frac{1}{L}\sum_{i=1}^{L}\left|\sum_{m=1}^M (h_m-1)s_m[i]\right|^2 + \frac{N_0}{T},
\end{eqnarray*}
where (a) follows because $s_m[i]$ is independent of the noise term $z[i]$; the last equality follows since the noise terms are i.i.d. for different $i$.

Defining $\bm{h}=[h_1,\allowbreak h_2,...,\allowbreak h_M]^\top$ and $\bm{s[i]}=[s_1[i],\allowbreak  s_2[i],\allowbreak  ...,\allowbreak  s_M[i]]^\top$, $\text{MSE}_\text{ML}$ can be written in a more compact form as
\begin{eqnarray}\label{eq:MSENaiveh}
\text{MSE}_\text{ML} \hspace{-0.2cm}&=&\hspace{-0.2cm} \frac{1}{L}\sum_{i=1}^{L}\left|(\bm{h-1})^\top\bm{s[i]}\right|^2 + \frac{N_0}{T} \nonumber\\
\hspace{-0.2cm}&=&\hspace{-0.2cm}  (\bm{h-1})^H \frac{1}{L}\sum_{i=1}^{L}\bm{s^*[i]s^\top[i]} (\bm{h-1}) + \frac{N_0}{T} \nonumber\\
\hspace{-0.2cm}&\triangleq&\hspace{-0.2cm}  (\bm{h-1})^H \bm{V} (\bm{h-1}) + \frac{N_0}{T},
\end{eqnarray}
in which $\bm{1}$ is an $M\times 1$ all-ones vector. In particular, the matrix $\bm{V}$ can be written as
\begin{eqnarray}\label{eq:V}
\bm{V}\triangleq
\begin{bmatrix}
\widehat{\mathbb{V}}_1 & \widehat{\mathbb{E}}^*_1\widehat{\mathbb{E}}_2 & \cdots & \widehat{\mathbb{E}}^*_1\widehat{\mathbb{E}}_M \\
\widehat{\mathbb{E}}^*_2\widehat{\mathbb{E}}_1 & \widehat{\mathbb{V}}_2 & \cdots & \widehat{\mathbb{E}}^*_2\widehat{\mathbb{E}}_M \\
\cdots & \cdots & \cdots & \cdots \\
\widehat{\mathbb{E}}^*_M\widehat{\mathbb{E}}_1 & \widehat{\mathbb{E}}^*_M\widehat{\mathbb{E}}_2 & \cdots & \widehat{\mathbb{V}}_M
\end{bmatrix},
\end{eqnarray}
where
\begin{eqnarray}\label{eq:sampleMean}
\widehat{\mathbb{E}}_m \triangleq \frac{1}{L}\sum_{i=1}^{L} s_m[i],~~
\widehat{\mathbb{V}}_m \triangleq \frac{1}{L}\sum_{i=1}^{L} |s_m[i]|^2.
\end{eqnarray}
are, respectively, the first and second sample moments of the symbols transmitted by the $m$-th device in one transmission.

In the aligned OAC with neither channel-gain nor time misalignment, it is straightforward from \eqref{eq:MSENaiveh}  that the MSE of the ML estimator is
\begin{eqnarray}\label{eq:MSENaive}
\text{MSE}_\text{ML} = \frac{N_0}{T}.
\end{eqnarray}
In other words, the MSE of the ML estimator is simply the noise variance in the aligned OAC. On the other hand, when there is channel-gain misalignment (i.e., the synchronous OAC), the additional MSE introduced by the channel-gain misalignment is $(\bm{h-1})^H \bm{V(h-1)}$.

\subsection{LMMSE estimation for the synchronous OAC}
According to \eqref{eq:MSENaiveh}, the MSE performance of the ML estimator can be poor when either $|\bm{h-1}|^2$ or the noise variance is large.
To improve the reconstruction performance, we put forth a Bayesian approach by letting each device transmit two pieces of information to the fusion center, whereby an LMMSE estimator can be devised.

Specifically, in each transmission, we let each device transmit the first sample moment $\widehat{\mathbb{E}}_m$ and the second sample moment $\widehat{\mathbb{V}}_m$ to the fusion center reliably (in a digital manner, with channel coding and automatic repeat request, for example) before the data transmission.
The fusion center then constructs a vector $\bm{\widehat{\mu}}$ and a matrix $\bm{D}$ from the first and second sample moments by
\begin{eqnarray}
\label{eq:mu}
\bm{\widehat{\mu}} = \left[\widehat{\mathbb{E}}_1,\widehat{\mathbb{E}}_2,\cdots,\widehat{\mathbb{E}}_M \right]^\top, \\
\label{eq:D}
\bm{D} = \text{diag}\left(\widehat{\mathbb{D}}_1,\widehat{\mathbb{D}}_2,\cdots,\widehat{\mathbb{D}}_M\right),
\end{eqnarray}
where $\widehat{\mathbb{D}}_m$ is defined to be the sample variance of the symbols transmitted by the $m$-th device in one transmission, i.e.,
\begin{eqnarray}
\widehat{\mathbb{D}}_m \triangleq \frac{1}{L}\sum_{i=1}^{L} |s_m[i]-\widehat{\mathbb{E}}_m|^2 = \widehat{\mathbb{V}}_m - |\widehat{\mathbb{E}}_m|^2.
\end{eqnarray}

Given $\bm{\widehat{\mu}}$ and $\bm{D}$, we now devise the LMMSE estimator.

\begin{lem}[positive definiteness of $\bm{D}$ and $\bm{V}$]
Matrix $\bm{D}$ is positive definite; matrix $\bm{V}$ is Hermitian positive definite.
\end{lem}

\begin{NewProof}
For the diagonal matrix $\bm{D}$, element $\widehat{\mathbb{D}}_m$ is the sample variance of the symbols transmitted by device $m$. This suggests that $\widehat{\mathbb{D}}_m>0$, $\forall m$. As a result, $\bm{D}$ is positive definite.

For the complex matrix $\bm{V}$, it can be seen from the definition that it is Hermitian, i.e., $\bm{V}^H=\bm{V}$. For any complex vector $\bm{x}=[x_1,x_2,...,x_M]\in\mathbb{C}^M$, $\bm{x\neq 0}$, we have
\begin{eqnarray*}
\hspace{-0.7cm}&&\bm{x}^H\bm{Vx} = \\
\hspace{-0.7cm}&&[x^*_1, x^*_2, ..., x^*_M]\frac{1}{L}\sum_{i=1}^{L}
\begin{bmatrix}
\begin{smallmatrix}
\left|s_1[i]\right|^2 & s^*_1[i]s_2[i] & \cdots & s^*_1[i]s_M[i] \\
s^*_2[i]s_1[i] & \left|s_2[i]\right|^2 & \cdots & s^*_2[i]s_M[i] \\
\cdots & \cdots & \cdots & \cdots \\
s^*_M[i]s_1[i] & s^*_M[i]s_2[i] & \cdots & \left|s_M[i]\right|^2
\end{smallmatrix}
\end{bmatrix}
\begin{bmatrix}
\begin{smallmatrix}
x^*_1 \\
x^*_2 \\
... \\
x^*_M
\end{smallmatrix}
\end{bmatrix} \\
\hspace{-0.7cm}&&
=\frac{1}{L}\sum_{i=1}^{L}\left|\sum_{m=1}^M x_ms_m[i]\right|^2 > 0.
\end{eqnarray*}
Thus, $\bm{V}$ is Hermitian positive definite.
\end{NewProof}

\begin{thm}[LMMSE estimation for the synchronous OAC]\label{thm:Syn_LMMSE}
Given a sequence of received samples $\bm{r}\in\mathbb{C}^L$ in \eqref{eq:III1}, an LMMSE estimator estimates the sequence $\bm{s}_+\in\mathbb{C}^L$ symbol-by-symbol by
\begin{eqnarray}\label{eq:LMMSEEstimator}
\widehat{s}_+[i] = \frac{\bm{h}^H\bm{D1}}{\bm{h}^H\bm{Dh}+\frac{N_0}{T}}r[i] + \left(\bm{1}\!-\!\frac{\bm{h}^H\bm{D1}}{\bm{h}^H\bm{Dh}+\frac{N_0}{T}}\bm{h}\right)^\top \!\!\bm{\widehat{\mu}}.
\end{eqnarray}
The MSE of the LMMSE estimator is
\begin{eqnarray}\label{eq:MSELMMSEh}
\text{MSE}_\text{LMMSE}= \bm{1}^\top\bm{D1}-\frac{\left|\bm{h}^H\bm{D1}\right|^2}{\bm{h}^H\bm{Dh}+\frac{N_0}{T}},
\end{eqnarray}
and we have $\text{MSE}_\text{LMMSE}\leq \text{MSE}_\text{ML}$.
\end{thm}

\begin{NewProof}
See Appendix~\ref{sec:AppA}.
\end{NewProof}

The LMMSE estimator can also be used in the aligned OAC. Considering the samples in \eqref{eq:II7}, the LMMSE estimator estimates the sequence $\bm{s}_+\in\mathcal{C}^L$ symbol-by-symbol by
\begin{eqnarray}\label{eq:LMMSEEstimatorAligned}
\widehat{s}_+[i] = \frac{\bm{1}^\top\bm{D1}}{\bm{1}^\top\bm{D1}+\frac{N_0}{T}}r[i] + \frac{\frac{N_0}{T}}{\bm{1}^\top\bm{D1}+\frac{N_0}{T}} \bm{1}^\top \!\!\bm{\widehat{\mu}}.
\end{eqnarray}
The MSE of the LMMSE estimator is
\begin{eqnarray}\label{eq:MSELMMSE}
\text{MSE}_\text{LMMSE}= \frac{ \frac{N_0}{T}\bm{1}^\top\bm{D1} }{\bm{1}^\top\bm{D1}+\frac{N_0}{T}}.
\end{eqnarray}
In particular, we have $\text{MSE}_\text{LMMSE}< \text{MSE}_\text{ML}$ since
\begin{eqnarray}\label{eq:MSEratio}
\frac{\text{MSE}_\text{ML}}{\text{MSE}_\text{LMMSE}}= \frac{\bm{1}^\top\bm{D1}+\frac{N_0}{T}}{\bm{1}^\top\bm{D1}} > 1.
\end{eqnarray}

To summarize, the ML estimator suffers from both channel-gain misalignment $\bm{h}$ and noise, whereas the LMMSE estimator can alleviate the estimation errors caused by both the channel-gain misalignment and noise, thanks to the prior information transmitted from the edge devices. The MSE gains are not analytically straightforward from \eqref{eq:MSENaiveh} and \eqref{eq:MSELMMSEh} for the general channel-gain misalignment $\bm{h}$. Therefore, let us consider the aligned OAC where $\bm{h=1}$. It can be seen from \eqref{eq:MSEratio} that the MSE gains of the LMMSE estimator over the ML estimator are $3$ dB when $\bm{1}^\top\bm{D1}= \frac{N_0}{T}$. When $\bm{1}^\top\bm{D1}\leq \frac{N_0}{T}$, the gains can be significant.

%
%
%
%
%
%
%
%

\section{Asynchronous OAC}\label{sec:IV}
In this section, we consider the most challenging case with both channel-gain and time misalignments in the received signal, which we refer to as the asynchronous OAC, and study how to devise the Bayesian estimators, leveraging the first and second sample moments from the edge devices.
To start with, we reproduce the received signal $r(t)$ in \eqref{eq:II5} below.
\begin{eqnarray}\label{eq:rt}
r(t)= \sum_{\ell=1}^L \sum_{m=1}^M {h}_m s_m[\ell]p(t-\tau_m-\ell T) + z(t),
\end{eqnarray}
where $s_m[\ell]$ is the $\ell$-th complex number transmitted from the $m$-th device; $h_m=|h_m|e^{j\phi_m}$ is the residual channel gain; $\tau_m$ is the time offset of the $m$-th device relative to the first device.

\subsection{Whitened Matched Filtering and Sampling}\label{sec:IVA}
As shown in Fig.~\ref{fig:2}, the symbols from different devices in the asynchronous OAC are misaligned in time.
If we follow standard signal processing flows in digital communications to process the received signal, i.e., matched filtering $r(t)$ by the pulse $p(t)$ and oversampling at $\{iT+\tau_k:i=1,2,...,L;k=1,2,...,M\}$ to collect sufficient statistics, the resulting samples would exhibit colored noise, which is much undesired in stochastic inference.
To circumvent this problem and obtain whitened samples from the received signal, we employ a bank of $M$ matched filters of different lengths to collect power judiciously from $r(t)$.\footnote{To illustrate our Bayesian approach, this paper considers the misaligned OAC with the rectangular pulse. When other pulses are used, the whitened matched filter can be designed accordingly. Take the root-raised-cosine pulse for example. A design of the whitened matched filter is given in \cite{PNC}.}
\begin{figure}[t]
  \centering
  \includegraphics[width=0.95\columnwidth]{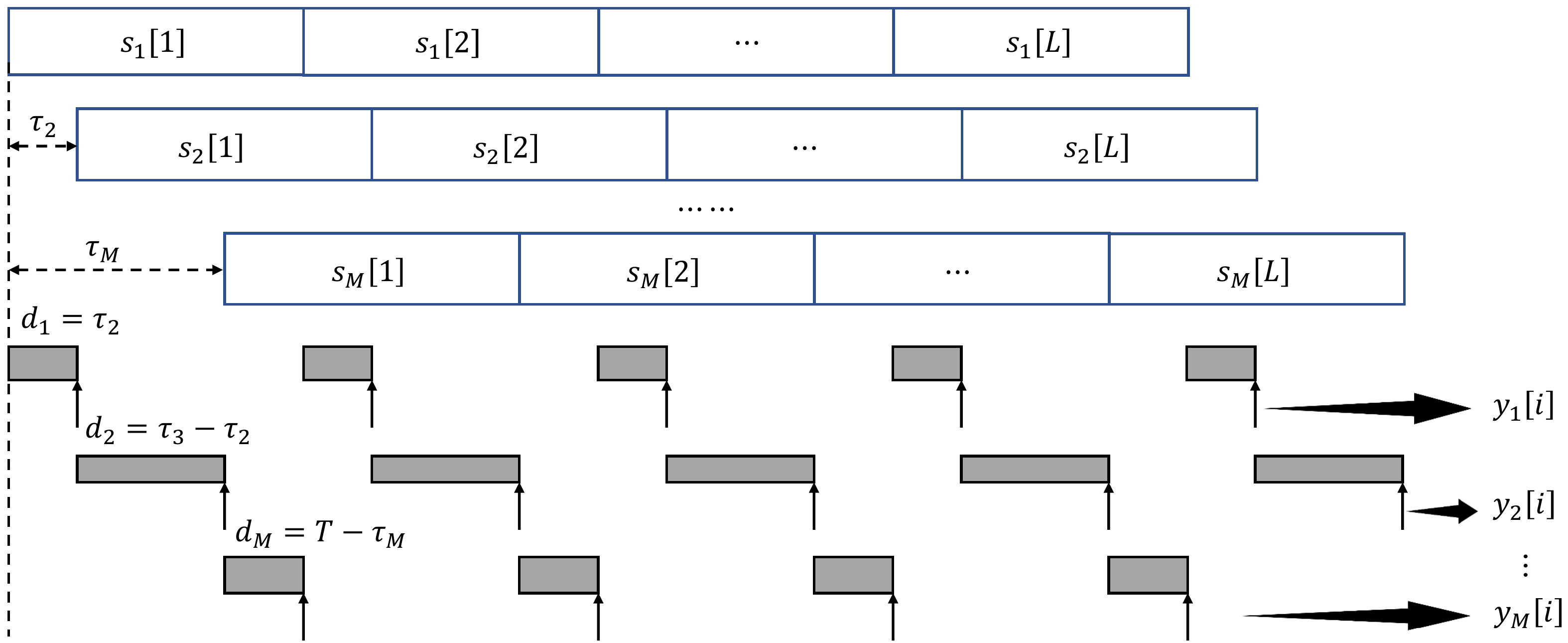}\\
  \caption{Matched filtering the received signal by a bank of $M$ filters of lengths $d_k=\tau_{k+1}-\tau_k$.}
\label{fig:3}
\end{figure}

The matched filtering and sampling processes are illustrated in Fig.~\ref{fig:3}. Specifically, the $M$ matched filters $\{p_k^\prime (t):k=1,2,...,M\}$ are defined as
\begin{eqnarray}\label{eq:matchedfilters}
p_k^\prime(t)=\frac{1}{2}\big[\text{sgn}(t+T)-\text{sgn}(t+T-d_k)  \big],
\end{eqnarray}
where the length of the $k$-th matched filter is $d_k=\tau_{k+1}-\tau_k$, $k=1,2,...,M$. For completeness, we define $\tau_{M+1}=T$.

The signal filtered by the $k$-th matched filter is given by
\begin{eqnarray*}
\hspace{-0.7cm}&& y_k(t)=\frac{1}{d_k}\int_{-\infty}^{\infty}r(\zeta)p^\prime_k(t-\zeta) \,d\zeta \\
\hspace{-0.7cm}&& =\!\frac{1}{d_k}\!\int_{-\infty}^{\infty}\!\!\! \left(\!\sum_{\ell=1}^L \!\sum_{m=1}^M\! {h}_m s_m[\ell]p(\zeta\!-\!\tau_m\!-\!\ell T) \!+\! z(\zeta)\!\right) \!p^\prime_k(t\!-\!\zeta) d\zeta,
\end{eqnarray*}
and we sample $y_k(t)$ at $(i-1)T\allowbreak+\tau_{k+1}:\allowbreak i=1,2,...,L,L+1$. The samples we get are
\begin{eqnarray}\label{eq:samples}
y_k[i]
\hspace{-0.2cm}&=&\hspace{-0.2cm} y_k(t=(i-1)T+\tau_{k+1})=\frac{1}{d_k} \int_{(i-1)T+\tau_{k}}^{(i-1)T+\tau_{k+1}} \nonumber\\
\hspace{-0.2cm}&& \sum_{m=1}^M h_m s_m[i-\mathbbm{1}_{m>k}]\,d\zeta +\frac{1}{d_k} \int_{(i-1)T+\tau_{k}}^{(i-1)T+\tau_{k+1}} \hspace{-0.3cm}z(\zeta)\,d\zeta  \nonumber\\
\hspace{-0.2cm}&=&\hspace{-0.2cm} \sum_{m=1}^M h_m s_m[i-\mathbbm{1}_{m>k}] + z_k[i],
\end{eqnarray}
where we have defined $s_m[0]=s_m[L+1]=0$, $\forall m$, for completeness.

Eq. \eqref{eq:samples} is very informative. Let us take a closer look:
\begin{enumerate}
\item Each sample in \eqref{eq:samples} is related to $M$ complex symbols, each of which comes from a different device.
\item It can be verified that $\mathbb{E}\left[ z_k[i] z_{k^\prime}[i^\prime]\right]=\frac{N_0 T}{d_k} \allowbreak\delta\big((i-i^\prime)(k-k^\prime)\big)$. This means that the noise sequence $\bm{z_k}$ is white: $z_k[i]\sim\mathcal{CN}(0, N_0 T/d_k)$ and is independent for different $k$ and $i$.
\item The symbols are aligned in time within the integral interval of the $M$-th matched filter. Specifically, let $k = M$, we have
\begin{eqnarray}\label{eq:samplesMth}
y_M[i]=\sum_{m=1}^M h_ms_m[i]+z_M[i],
\end{eqnarray}
where $z_M[i]\sim\mathcal{CN}(0,N_0 T/d_M)$. Unlike the outputs of other matched filters, the sampling outputs of the $M$-th matched filter form a synchronous OAC with the noise variance being amplified by $T/d_M$ times.
\end{enumerate}


The third observation suggests that the ML and LMMSE estimators designed for the synchronous OAC can also be used in the asynchronous case, utilizing the outputs of the $M$-th matched filter only (for the purpose of differentiation, we add a prefix ``p-'' before the ML and LMMSE estimators since only partial samples are used here).

\begin{cor}[MSEs of the p-ML and p-LMMSE estimators in asy\-nch\-ronous OAC]\label{thm:Asyn_naiveLMMSE}
In asynchronous OAC, given the output of the $M$-th matched filter in \eqref{eq:samplesMth}, the MSEs of the ML and LMMSE estimators are
\begin{eqnarray}\label{eq:MSENaiveAsyn}
\text{MSE}_\text{p-ML} \hspace{-0.2cm}&=&\hspace{-0.2cm} (\bm{h-1})^H \bm{V} (\bm{h-1}) +\frac{N_0 T}{d_M}, \\
\label{eq:MSELMMSEAsyn}
\text{MSE}_\text{p-LMMSE} \hspace{-0.2cm}&=&\hspace{-0.2cm} \bm{1}^\top\bm{D1}-\frac{\left|\bm{h}^H\bm{D1}\right|^2}{\bm{h}^H\bm{Dh}+\frac{N_0 T}{d_M}}.
\end{eqnarray}
\end{cor}

From \eqref{eq:MSENaiveAsyn} and \eqref{eq:MSELMMSEAsyn}, it is clear that the MSEs of the p-ML and p-LMMSE estimators hinge on the maximum time offset $\tau_M$ as it determines the duration of the $M$-th matched filter $d_M=T-\tau_M$. Take the p-LMMSE estimator for example. In the synchronous OAC, we have $\tau_M=0$, and hence, $d_M=T$. In the asynchronous OAC, on the other hand, $\text{MSE}_\text{p-LMMSE}$ increases with $\tau_M$.
To the extent that as $\tau_M\allowbreak\to\allowbreak T$ (hence $d_M\allowbreak \to\allowbreak 0$), $\text{MSE}_\text{p-LMMSE}\allowbreak\to\allowbreak \bm{1}^\top\bm{D1}$.
In the next section, we shall devise more powerful estimators that make use of the samples from all matched filters.

\subsection{ML estimation versus LMMSE estimation}\label{sec:IVB}
To start with, let us rewrite all the samples given by \eqref{eq:samples} into a more compact form as
\begin{eqnarray}\label{eq:samplesMat}
\bm{y}= \bm{Gs+z},
\end{eqnarray}
where $\bm{y}$, $\bm{s}$, and $\bm{z}$ are defined as
\begin{small}
\begin{eqnarray*}
&&\hspace{-0.7cm}  \bm{y}\triangleq \Big[y_1[1],y_2[1],...,y_M[1],y_1[2],y_2[2],...,y_M[2],..., \\
&&\hspace{-0.7cm}  y_1[L],y_2[L],...,y_M[L], y_1[L\!+\!1],y_2[L\!+\!1],...,y_{M\!-\!1}[L\!+\!1] \Big]^\top, \\
&&\hspace{-0.7cm}  \bm{s}\triangleq\Big[s_1[1],s_2[1],...,s_M[1],s_1[2],s_2[2],...,s_M[2],..., \\
&&\hspace{-0.7cm}  s_1[L],s_2[L],...,s_M[L] \Big]^\top, \\
&&\hspace{-0.7cm}  \bm{{z}}\triangleq\Big[{z}_1[1],{z}_2[1],...,{z}_M[1],{z}_1[2],{z}_2[2],...,{z}_M[2],..., \\
&&\hspace{-0.7cm}  {z}_1[L],{z}_2[L],...,{z}_M[L], {z}_1[L\!+\!1],{z}_2[L\!+\!1],...,{z}_{M\!-\!1}[L\!+\!1] \Big]^\top,
\end{eqnarray*}
\end{small}
and the $M(L+1)-1$ by $ML$ coefficient matrix $\bm{G}$ is
\begin{eqnarray*}
\bm{G}\triangleq
\begin{bmatrix}
\begin{smallmatrix}
h_1 &            &       &             &             &               &    &               &    &     \\
h_1 & h_2 &       &             &             &               &    &               &    &     \\
...        & h_2 & ...   &             &             &               &    &               &     &   \\
h_1 & ...        & ...   &  h_M &             &               &    &               &    &    \\
           & h_2 & ...   &  h_M & h_1  &               &    &               &    &    \\
           &            & ...   &  ...        & h_1  & h_2    &    &               &     &   \\
           &            &       &  h_M & ...         & h_2    & ...  &             &    &    \\
           &            &       &             & h_1  & ...           & ...  & h_M  &    &    \\
           &            &       &             &             & h_2    & ...  & h_M  & ...&       \\
           &            &       &             &             &               & ...  & ...         & ... &   \\
           &            &       &             &             &               &      & h_M  & ... &     \\
           &            &       &             &             &               &      &             & ... &
\end{smallmatrix}
\end{bmatrix}.
\end{eqnarray*}

Eq.~\eqref{eq:samplesMat} is in the form of a classic inter-symbol interference (ISI) channel model in digital communications with two main differences:
\begin{enumerate}[leftmargin=0.45cm]
\item The sequence of transmitted symbols $\bm{s}$ are continuous complex values instead of discrete constellations. In digital communications, the discrete constellation is a kind of prior information to the receiver whereby the detection space is naturally narrowed down to the possible constellation points only. In OAC, however, we do not have such prior information due to the continuous nature of the transmitted signal. The estimation space is thus infinitely large.
\item Our aim is not to estimate the transmitted symbols $\bm{s}$, but a linear transformation of $\bm{s}$:
\begin{eqnarray}\label{eq:s_plus_matrix}
\bm{s_+}= \bm{F}\bm{s},
\end{eqnarray}
where $L$ by $ML$ dimensional matrix $\bm{F}$ is given by
\begin{eqnarray*}
\bm{F} = \begin{bmatrix}
\bm{1}^\top &                    &                    &                    \\
                   & \bm{1}^\top &                    &                    \\
                   &                    & ...                &                    \\
                   &                    &                    & \bm{1}^\top
\end{bmatrix},
\end{eqnarray*}
in which $\bm{1}^\top$ is a $1\times M$ all-ones vector.
\end{enumerate}

To estimate $\bm{s}_+$ in \eqref{eq:s_plus_matrix}, a viable estimator that utilizes all the samples $\bm{y}$ is the ML estimator.

\begin{defi}[ML estimation for the asynchronous OAC]\label{defi:2}
Given a sequence of samples $\bm{y}\in\mathcal{C}^{M(L+1)}$ in \eqref{eq:samplesMat}, the ML estimate of sequence $\bm{s}_+\in\mathcal{C}^L$ is
\begin{eqnarray}\label{eq:ML}
\widehat{\bm{s}}^\text{ml}_+ = \bm{F}(\bm{G^H}\bm{\Sigma^{-1}_{{z}} G})^{-1}\bm{G^H\Sigma^{-1}_{{z}}y},
\end{eqnarray}
where $\bm{\Sigma_{z}}$ denotes the covariance matrix of the noise sequence $\bm{z}$. In particular, $\bm{\Sigma_{z}}$ is a diagonal matrix since $\bm{z}$ is white.
\end{defi}

The MSE of the ML estimator can be derived as
\begin{eqnarray*}
\text{MSE} \hspace{-0.3cm}&=&\hspace{-0.3cm} \frac{1}{L}\mathbb{E}\!\left[(\widehat{\bm{s}}^\text{ml}_+\!\!-\!\!{\bm{s}}_+)^H(\widehat{\bm{s}}^\text{ml}_+\!\!-\!\!{\bm{s}}_+)\right]
\!=\!\!\frac{1}{L}\text{Tr}\big\{\bm{F}(\bm{G^H}\bm{\Sigma^{-1}_{{z}} G})^{-1} \\
&&\quad \bm{G^H\Sigma^{-1}_{z}\mathbb{E}[\bm{z}}\bm{z}^H ] \bm{\Sigma}^{-H}_{\bm{z}}\bm{G}(\bm{G}^H\bm{\Sigma^{-1}_{{z}} G})^{-H}\bm{F}^H \big\}\\
\hspace{-0.3cm}&=&\hspace{-0.3cm} \frac{1}{L}\text{Tr}\big\{\bm{F}(\bm{G}^H\bm{\Sigma^{-1}_{{z}} G})^{-H}\bm{F}^H \big\} \\
\hspace{-0.3cm}&=&\hspace{-0.3cm} \frac{1}{L}\text{Tr}\big\{\bm{F}(\bm{G}^H\bm{\Sigma^{-1}_{{z}} G})^{-1}\bm{F}^\top \big\}.
\end{eqnarray*}

It is revealed in our prior work \cite{techFL} that ML estimation is much susceptible to noise due to the infinitely large estimation space. As a result, error propagation and noise enhancement are severe with the ML estimator.

Unlike ML estimation, which treats the transmitted sequence $\bm{s}$ as a constant sequence, this paper treats $\bm{s}$ as a random sequence and leverages a Bayesian approach to address the problems faced by ML estimation.
As in the synchronous case, we will show that the estimation performance can be significantly improved by making good use of two pieces of information (i.e., the first and second sample moments) transmitted from each edge device.


\begin{thm}[LMMSE estimation for the asynchronous OAC]\label{thm:prop7}
Given a sequence of samples $\bm{y}\in\mathcal{C}^{M(L+1)}$ in \eqref{eq:samplesMat}, an LMMSE estimator estimates the sequence $\bm{s}_+\in\mathcal{C}^L$ by
\begin{eqnarray}\label{eq:LMMSE_misaligned}
\widehat{\bm{s}}^\text{LMMSE}_+ = \bm{Ay}+(\bm{F-AG})\widetilde{\bm{\mu}},
\end{eqnarray}
where $\bm{A}\triangleq\bm{F\widetilde{D}G}^H(\bm{G\widetilde{D}G}^H+\bm{\Sigma_z})^{-1}$. The vector $\widetilde{\bm{\mu}}\triangleq\mathbb{E}[\bm{s}]$ and matrix $\bm{\widetilde{D}}\triangleq\mathbb{E}[\bm{ss}^H]-\mathbb{E}[\bm{s}]\mathbb{E}^H[\bm{s}]$ can be constructed from the first and second sample moments transmitted from the devices, giving
\begin{eqnarray*}
\widetilde{\bm{\mu}} \hspace{-0.2cm}&=&\hspace{-0.2cm} \left[\widehat{\mathbb{E}}_1,\widehat{\mathbb{E}}_2,...,\widehat{\mathbb{E}}_M, \widehat{\mathbb{E}}_1,\widehat{\mathbb{E}}_2,...,\widehat{\mathbb{E}}_M, ...,\widehat{\mathbb{E}}_1,\widehat{\mathbb{E}}_2,...,\widehat{\mathbb{E}}_M \right]^\top, \\
\bm{\widetilde{D}}
\hspace{-0.2cm}&=&\hspace{-0.2cm}
\text{diag}\left(\widehat{\mathbb{D}}_1,...,\widehat{\mathbb{D}}_M,\widehat{\mathbb{D}}_1,...,\widehat{\mathbb{D}}_M,...,\widehat{\mathbb{D}}_1,...,\widehat{\mathbb{D}}_M\right).
\end{eqnarray*}

The MSE of the LMMSE estimator is
\begin{eqnarray}
\text{MSE}_\text{LMMSE}\!=\! \frac{1}{L}\text{Tr}\!\left[(\bm{AG\!\!-\!\!F})\widetilde{\bm{D}}(\bm{AG\!\!-\!\!F})^H \!\!+\! \bm{A\Sigma_zA}^H \right].
\end{eqnarray}
\end{thm}

\begin{NewProof}
See Appendix~\ref{sec:AppA2}.
\end{NewProof}
%

Leveraging the first and second sample moments from the edge devices as a kind of prior information, the LMMSE estimator is able to reduce the estimation space by a large margin, and hence, improve the MSE performance.

However, a problem with the LMMSE estimator is the high computational complexity since the dimensionalities of the matrices in \eqref{eq:LMMSE_misaligned} grow linearly with the number of devices $M$ and the packet length $L$. As a result, \eqref{eq:LMMSE_misaligned} is computationally expensive for large $M$ and $L$.
Let us assess the computational complexity of \eqref{eq:LMMSE_misaligned} by the matrix inversion
$(\bm{G\widetilde{D}G}^H+\bm{\Sigma_z})^{-1}$, which is the most computationally demanding part of \eqref{eq:LMMSE_misaligned}. To invert an $n$ by $n$ matrix, the best proven lower bound of the computational complexity is $\Omega(n^2\log n)$ \cite{complexity}. Thus, the complexity of \eqref{eq:LMMSE_misaligned} is $\Omega(M^2L^2\log(ML))$. In OAC systems, the packet length $L$ is often much larger than the number of devices $M$.
Let us fix $M$ as a constant, the computational complexity of \eqref{eq:LMMSE_misaligned} is then $\Omega(L^2\log L)$.

In the next subsection, we shall solve this problem by exploiting the sparsity of the coefficient matrix and developing a sum-product MAP (SP-MAP) estimator.
The MSE performance of the SP-MAP estimator is on an equal footing with the LMMSE estimator, but its computational complexity is only $\Omega(L)$.

\subsection{MAP Estimation}\label{sec:IVC}
\begin{figure*}[t]
  \centering
  \includegraphics[width=1.6\columnwidth]{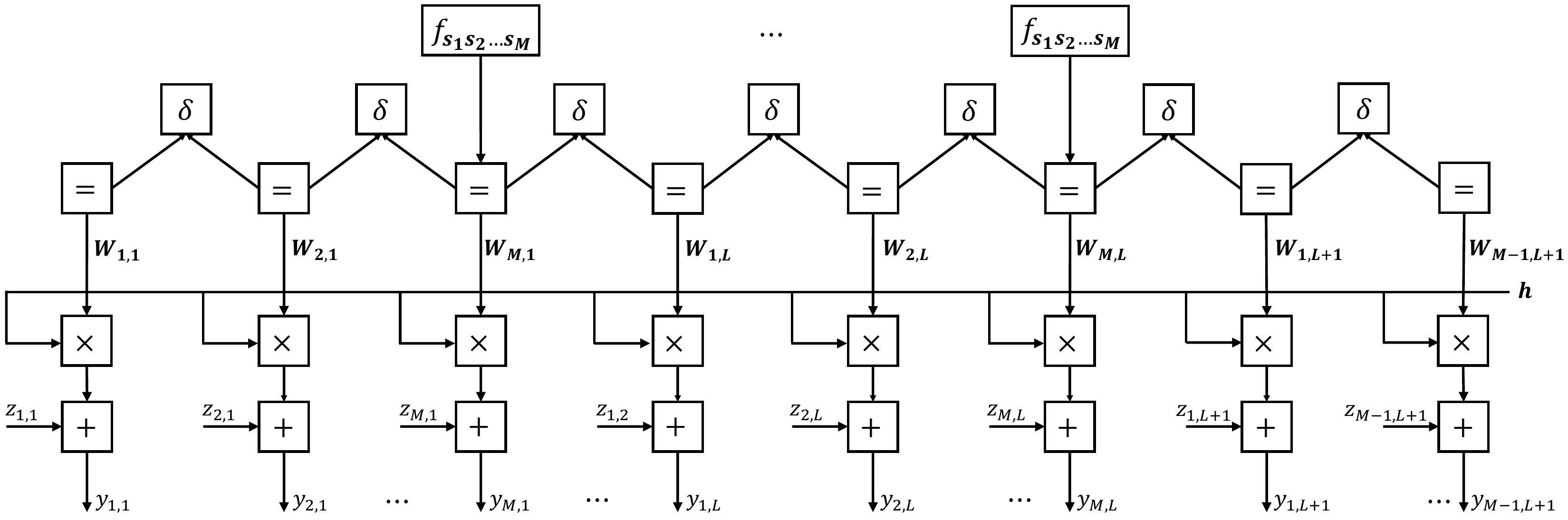}\\
  \caption{A graphical interpretation of the factorization in \eqref{eq:factorization}. The high-dimensional variable $\bm{W}_{k,i}=\mathcal{V}(y_k[i])=\allowbreak\{s_1[i],\allowbreak s_2[i],...,\allowbreak s_k[i],\allowbreak s_{k+1}[i-1],\allowbreak s_{k+2}[i-1],...,\allowbreak s_M[i-1]\}$. To simplify notations, we denote $y_k[i]$ and $z_k[i]$ by $y_{k,i}$ and $z_{k,i}$, respectively, in the figure. }
\label{fig:5}
\end{figure*}

To tackle the high complexity of the LMMSE estimation and to devise a practical estimator for the misaligned OAC, this subsection resorts to MAP estimation and puts forth an SP-MAP estimator for the asynchronous OAC.

\begin{defi}[MAP estimation for the asynchronous OAC]\label{defi:MAPEstiamtor}
Given the white samples $\{y_k[i]\}$ in \eqref{eq:samples}, an MAP estimator estimates the element of the target sequence $\bm{s}_+\in\mathcal{C}^L$ by
\begin{eqnarray}\label{eq:MAP}
\widehat{s}_+[i]=\arg\max_{s_+[i]}\Pr\left( {s}_+[i] = \sum_{m=1}^M s_m[i] ~\Big|~ \bm{y} \right),
\end{eqnarray}
where
\begin{eqnarray*}
\hspace{-0.5cm}&& \Pr\left(\!{s}_+[i]\!=\!\sum_{m=1}^M\!s_m[i]~\Big|~ \bm{y}\!\right)= \int_{\sum_{m=1}^Ms_m[i]={s}_+[i]} \\
\hspace{-0.5cm}&&  f(\!s_1[i],\!s_2[i],...,\!s_M[i] | \bm{y})d(\!s_1[i],\!s_2[i],...,\!s_M[i]).
\end{eqnarray*}
\end{defi}

As can be seen, to obtain the MAP estimate, a first step is to derive the joint posterior probability distribution $f(\!s_1[i],\allowbreak\!s_2[i],...,\allowbreak\!s_M[i] | \bm{y})$. For this purpose, we shall start from the joint posterior distribution of all transmitted symbols conditioned on the samples observed at the receiver, i.e., $f(\bm{s_1},\allowbreak\bm{s_2},...,\bm{s_M} | \bm{y})$.
To ease exposition, we name
$f(\!s_1[i],\allowbreak\!s_2[i],...,\!s_M[i] | \bm{y})$ the marginal posterior distribution and $f(\bm{s_1},\allowbreak\bm{s_2},...,\bm{s_M} | \bm{y})$ the global posterior distribution.

The global posterior distribution can be factorized in the following way:
\begin{eqnarray}\label{eq:factorization}
\hspace{-1.6cm}&& f(\bm{s_1},\bm{s_2},...,\bm{s_M} | \bm{y}) \nonumber\\
\hspace{-1.6cm}&& \propto f(\bm{y}|\bm{s_1},\bm{s_2},...,\bm{s_M})f(\bm{s_1},\bm{s_2},...,\bm{s_M}) \nonumber\\
\hspace{-1.6cm}&& \overset{(a)}{\propto} \prod_{k=1}^M \prod_{i=1}^{L+1} f(y_k[i]|\bm{s_1},\bm{s_2},...,\bm{s_M})\prod_{m=1}^{M} f(\bm{s_m}) \nonumber\\
\hspace{-1.6cm}&& \overset{(b)}{\propto} \prod_{k=1}^M \prod_{i=1}^{L+1} f(y_k[i]|\mathcal{V}(y_k[i]))\prod_{m=1}^{M} \prod_{i=1}^{L}f(s_m[i]),
\end{eqnarray}
where $\propto$ stands for ``proportional to''. Step (a) follows because i) given the transmitted sequence $\bm{s_1}$, $\bm{s_2}$, ..., $\bm{s_M}$, all the samples $\bm{y}$ are independent since the noise sequence is white; ii) the transmitted symbols $\bm{s_m}$ from different devices are independent. Step (b) follows since
\begin{enumerate}[label=\roman*),leftmargin=0.42cm]
\item Each sample $y_k[i]$ is only related to a set of complex symbols $\mathcal{V}(y_k [i])\allowbreak=\{s_1[i],\allowbreak s_2[i],...,s_k [i],s_{k+1}[i-1],s_{k+2}[i-1],\allowbreak...,s_M [i-1]\}$. We call them the {\it neighbor symbols} of the sample $y_k[i]$. In particular, the number of non-zero symbols in $\mathcal{V}(y_k[i])$ is
\begin{eqnarray}\label{eq:neighbors}
\left|\mathcal{V}(y_k[i])\right| =
\begin{cases}
k, & \text{when}~i=1; \\
M, & \text{when}~1\leq i\leq L; \\
M-k, &  \text{when}~i=L+1.
\end{cases}
\end{eqnarray}
\item $f(\bm{s_m})$ is the prior distribution of the transmitted symbols from the $m$-th device. To construct this information, we assume the symbols of each edge device are generated from a Gaussian distribution in an i.i.d. manner  at the fusion center. In particular, the Gaussian distribution is parameterized by the first and second sample moments transmitted from each edge device. This is a plausible assumption since there is no randomness once a packet is generated at the transmitter and the receiver can assume they are sampled from an i.i.d. Gaussian with mean and variance being its sample mean and sample variance. We emphasize that the devices have to generate the first and second sample moments for each new packet and transmit it to the receiver. The receiver then
    estimates different packets using different prior information.
\end{enumerate}

The factorizations in \eqref{eq:factorization} can be depicted by a graphical model \cite{kschischang2001factor,loeliger2007factor,PNC,wang2019gaussian}, as shown in Fig.~\ref{fig:5}, where we use a Forney-style factor graph \cite{loeliger2007factor} to represent the factorization.
Specifically, each edge in the graph corresponds to a variable in \eqref{eq:samples}, e.g., an observation $y_k[i]$ or a noise term $z_k[i]$. The variable $\bm{W}_{k,i}$ is a high-dimensional variable consisting of all complex symbols in $\mathcal{V}(y_k[i])$, i.e., $\bm{W}_{k,i}=\mathcal{V}(y_k[i])=\allowbreak\{s_1[i],\allowbreak s_2[i],...,\allowbreak s_k[i],\allowbreak s_{k+1}[i-1],\allowbreak s_{k+2}[i-1],...,\allowbreak s_M[i-1]\}$.

The equality function/constraint ``$=$'' in Fig.~\ref{fig:5} means that the variables connecting to this function are exactly the same (but may have different posterior distributions). The compatibility function $\delta$, on the other hand, represents the constraint that the values of the common symbols contained in the adjacent variables must be equal.
For example, $\bm{W}_{1,1}=\{s_1[1]\}$, $\bm{W}_{2,1}=\allowbreak \{s_1[1],\allowbreak s_2[1]\}$, and the common symbol between $\bm{W}_{1,1}$ and $\bm{W}_{2,1}$ is $s_1[1]$. Therefore, we have to add a constraint $\delta(\bm{W}_{1,1},\bm{W}_{2,1})$ between $\bm{W}_{1,1}$ and $\bm{W}_{2,1}$ to ensure that the values of $s_1[1]$ in $\bm{W}_{1,1}$ and $\bm{W}_{2,1}$ are the same. In general, for any two adjacent variables $\bm{W}$ and $\bm{W}^\prime$ connecting to the same delta function $\delta(\bm{W},\bm{W}^\prime)$, we have
\begin{equation*}
\delta(\bm{W},\bm{W}^\prime)=\begin{cases}
1, & \hspace{-0.2cm}\text{if the values of all common symbols}\\
&\hspace{-0.2cm}\text{between}~\bm{W}~\text{and}~\bm{W}^\prime~\text{are equal}; \\
0, & \hspace{-0.2cm}\text{otherwise.}
\end{cases}
\end{equation*}
Succinctly speaking, function $\delta$ is an on-off function ensuring that the messages passed from $\bm{W}$ to $\bm{W}^\prime$ and that passed from $\bm{W}^\prime$ to $\bm{W}$ satisfy the constraint that the values of the common symbols between $\bm{W}$ and $\bm{W}^\prime$ are equal.

Finally, the prior information $f_{\bm{s_1,s_2,...,s_M}}$ is an $M$-di\-me\-nsional Gaussian, the mean vector and covariance matrix of which are given in \eqref{eq:mu} and \eqref{eq:D}, respectively. To avoid unnecessary loops, the prior information is added every $M$ samples, as shown in Fig.~\ref{fig:5}.

\subsection{The SP-MAP Estimator}\label{sec:IVD}
The marginal posterior distribution $f(\allowbreak s_1[i],\allowbreak s_2[i],...,\allowbreak  s_M[i]|\allowbreak \bm{y})$ is a marginal function of the global posterior distribution $f(\bm{s_1},\allowbreak\bm{s_2},...,\bm{s_M}|\bm{y})$. Therefore, it can be derived by a marginalization process operated on Fig.~\ref{fig:5}, which can be implemented efficiently via the sum-product algorithm.

A caveat here is that, unlike digital communications, all the variables in Fig.~\ref{fig:5} are continuous random variables since the transmitted symbols $\{\bm{s_1},\bm{s_2},...,\bm{s_M}\}$ are continuous complex values.
Therefore, the messages to be passed on the graph are continuous probability density functions (PDFs) as opposed to discrete probability mass functions (PMFs).
In Theorem~\ref{thm:AMP} below, we point out an important result that  all the messages passed on the tree are multivariate Gaussian distributions.
This suggests that we can parameterize the PDFs by their mean vectors and covariance matrices -- passing these parameters is equivalent to passing the continuous PDFs.

\begin{algorithm}[t]
\caption{Analog message passing for SP-MAP estimation.}\label{algo:1}
\setlength{\tabcolsep}{1mm} 
\begin{algorithmic}[1]
\State{\textbf{Input:} Samples $\bm{y}$ and coefficient matrix $\bm{D}$.}
\State{\textbf{Output:} The marginal posterior distribution $f(\bm{s[i]}| \bm{y})$.}
\State{\# Initialization:}
\For{$k=1,2,...,M$ and $i=1,2,...,L+1$}
\State{$\bm{w_{k,i}}=\mathcal{V}(y_k[i])$;}
\State{Compute the information about $\bm{w_{k,i}}$ contained in each sample $y_k[i]$, i.e., $f_b(\bm{w_{k,i}})$, following \eqref{eq:fb}.}
\EndFor
\State{\# Forward message passing:}
\For{$i=1,2,...,L+1$}
\For{$k=1,2,...,M$}
\State{Compute the information about $\bm{w_{k,i}}$ contained in all the samples $\{y_{k^\prime}[i^\prime]:k^\prime<k,i^\prime<i\}$, i.e., $f_{\ell}(\bm{w_{k,i}})$, following \eqref{eq:fr} and \eqref{eq:fl_right}.}
\EndFor
\EndFor
\State{\# Backward message passing:}
\For{$i=L+1,L,...,2,1$}
\For{$k=M,M-1,...,2,1$}
\State{Compute the information about $\bm{w_{k,i}}$ contained in all the samples $\{y_{k^\prime}[i^\prime]:k^\prime>k,i^\prime>i\}$, i.e., $f^\prime_{r}(\bm{w_{k,i}})$, following \eqref{eq:backward1} and \eqref{eq:backward2}.}
\EndFor
\EndFor
\State{\# Marginalization:}
\For{$i=1,2,...,L$}
\State{Compute the marginal posterior distribution $f(\bm{s[i]}| \bm{y})$, as per \eqref{eq:marginal}.}
\EndFor
\end{algorithmic}
\end{algorithm}

\begin{thm}[Conditional Gaussian of the posterior distributions]\label{thm:AMP}
Consider the MAP estimator defined in Definition \ref{defi:MAPEstiamtor}. Let $\bm{s[i]}=(s_1[i],\allowbreak s_2[i], \allowbreak ..., \allowbreak s_M[i])$, we have the following results:
\begin{enumerate}[leftmargin=0.5cm]
\item The marginal posterior distribution $f(\bm{s[i]} |\allowbreak \bm{y})$ is an $M$-dimensional complex Gaussian distribution, giving
\begin{eqnarray}\label{eq:Marginal}
\hspace{-0.7cm}&& f(\bm{s[i]}|\bm{y})\sim \\
\hspace{-0.7cm}&& \mathcal{N}\left(\bm{s[i]},
\bm{\mu_{s[i]}}=
\begin{bmatrix}
\bm{\mu^\mathfrak{r}_{s[i]}} \\
\bm{\mu^\mathfrak{i}_{s[i]}}
\end{bmatrix},
\bm{\Sigma_{s[i]}} =
\begin{bmatrix}
\bm{\Sigma^{\mathfrak{rr}}_{s[i]}} & \bm{\Sigma^{\mathfrak{ri}}_{s[i]}} \nonumber\\
\bm{\Sigma^{\mathfrak{ir}}_{s[i]}} & \bm{\Sigma^{\mathfrak{ii}}_{s[i]}}
\end{bmatrix}
\right),
\end{eqnarray}
where $\bm{\mu_{s[i]}}$  is a $2M$ by $1$ real vector consisting of the real and imaginary parts of the mean of $\bm{s[i]}$, that is, $\bm{\mu^\mathfrak{r}_{s[i]}}$ and $\bm{\mu^\mathfrak{i}_{s[i]}}$ are the real and imaginary parts of sequence $\mathbb{E}[\bm{s[i]}]^\top$; the matrix $\bm{\Sigma_{s[i]}}$ is a $2M$ by $2M$ covariance matrix. The moment parameters $(\bm{\mu_{s[i]}},\bm{\Sigma_{s[i]}})$ can be computed by an analog message passing process described in Algorithm~\ref{algo:1}.
\item The posterior distribution $f({s}_+[i]=\allowbreak\sum_{m=1}^M s_m[i]| \allowbreak\bm{y})$ is a complex Gaussian distribution, giving
\begin{eqnarray}\label{eq:splus}
\hspace{-0.7cm}&& f({s}_+[i]| \allowbreak\bm{y})\sim \\
\hspace{-0.7cm}&& \mathcal{N}\left(
{s}_+[i],
\bm{\mu}_{s_+[i]}\!=\!
\begin{bmatrix}
\mu^\mathfrak{r}_{s_+[i]} \\
\mu^\mathfrak{i}_{s_+[i]}
\end{bmatrix},
\bm{\Sigma}_{s_+[i]} \!=\!
\begin{bmatrix}
\Sigma^{\mathfrak{rr}}_{s_+[i]} & \Sigma^{\mathfrak{ri}}_{s_+[i]} \nonumber\\
\Sigma^{\mathfrak{ir}}_{s_+[i]} & \Sigma^{\mathfrak{ii}}_{s_+[i]}
\end{bmatrix}
\right),
\end{eqnarray}
where $\mu^\mathfrak{r}_{s_+[i]} = \bm{1}^\top\bm{\mu^\mathfrak{r}_{s[i]}}$,
$\mu^\mathfrak{i}_{s_+[i]} = \bm{1}^\top\bm{\mu^\mathfrak{i}_{s[i]}}$,
$\Sigma^{\mathfrak{rr}}_{s_+[i]} = \bm{1}^\top \allowbreak \bm{\Sigma^{\mathfrak{rr}}_{s[i]}} \allowbreak \bm{1}$,
$\Sigma^{\mathfrak{ri}}_{s_+[i]} = \bm{1}^\top \allowbreak \bm{\Sigma^{\mathfrak{ri}}_{s[i]}} \allowbreak \bm{1}$,
$\Sigma^{\mathfrak{ir}}_{s_+[i]} = \bm{1}^\top \allowbreak \bm{\Sigma^{\mathfrak{ir}}_{s[i]}} \allowbreak \bm{1}$,
$\Sigma^{\mathfrak{ii}}_{s_+[i]} = \bm{1}^\top \allowbreak \bm{\Sigma^{\mathfrak{ii}}_{s[i]}} \allowbreak \bm{1}$.
\end{enumerate}
\end{thm}

\begin{NewProof}
See Appendix~\ref{sec:AppB}.
\end{NewProof}

Based on Theorem~\ref{thm:AMP}, the MAP estimator in Definition~\ref{defi:MAPEstiamtor} can be refined as follows.

\begin{defi}[SP-MAP estimation for the asynchronous OAC]\label{defi:SP_MAP}
Given the output of the $M$ matched filters $\bm{y}$ in \eqref{eq:samplesMat}, an SP-MAP estimator first computes the moment parameters of the Gaussian distribution $f(\bm{s[i]} |\allowbreak \bm{y})$ in \eqref{eq:Marginal} by analog message passing. From the mean vector $\bm{\mu_{s[i]}}= [\bm{\mu^\mathfrak{r}_{s[i]}},\allowbreak\bm{\mu^\mathfrak{i}_{s[i]}}]^\top$, the SP-MAP estimator estimates $s_+[i]$ by
\begin{eqnarray}\label{eq:SPAMAP}
\widehat{s}_+[i] = \bm{1}^\top\bm{\mu^\mathfrak{r}_{s[i]}} +j \bm{1}^\top\bm{\mu^\mathfrak{i}_{s[i]}}.
\end{eqnarray}
\end{defi}

Eq.~\eqref{eq:SPAMAP} can be understood in the following way: since $f({s}_+[i]| \allowbreak\bm{y})$ is Gaussian, its mean vector maximizes the posterior probability.
As per the MAP rule in \eqref{eq:MAP}, an SP-MAP estimator chooses the mean of $f({s}_+[i]| \allowbreak \bm{y})$, i.e., \eqref{eq:SPAMAP}, as the MAP estimate.

Finally, we compare the complexity of the SP-MAP estimator against that of the LMMSE estimator.
With Gaussian message passing, the messages passed on the graph are simply the parameters of the Gaussian distributions instead of the continuous Gaussian PDFs.
Thus, the computations involved in the analog message passing are only 1) the sum of 2M-dimensional vectors/matrices, and 2) 2M-dimensional matrix inversion.
The computational complexity of the SP-MAP estimator is then $\Omega(LM^3\log M)$. If we fix $M$ as a constant, the computational complexity of the SP-MAP estimator is simply $\Omega(L)$. In contrast, the computational complexity of the LMMSE estimator is $\Omega(L^2\log L)$.

\begin{rem}
Compared with the LMMSE estimator, the SP-MAP estimator breaks the $ML \times ML$ matrix inversion into the inversion of $ML$ smaller matrices of dimension $2M \times 2M$; thus, significantly reducing the computational complexity. This is thanks to the sparsity of the matrix $\bm{G}$: each sample $y_k[i]$ is only related to a set of complex symbols $\mathcal{V}(y_k [i])$.
\end{rem}

\section{Numerical and Simulation Results}\label{sec:V}
This section evaluates the MSE performance of various Bayesian OAC estimators devised in this paper benchmarked against the ML estimator. Specifically, we consider a MAC where $M=4$ devices communicate with a fusion center via OAC. There can be channel-gain and time misalignments among the received signal. We shall evaluate the MSEs of different estimators under various degrees of channel-gain misalignment, time misalignment, and EsN0:
\begin{enumerate}
\item The residual channel gain of the $m$-th device is $h_m=|h_m|e^{j\phi_m}$. In the simulations, we set $|h_m|=1$, $\forall m$, and focus on the impact of the phase misalignment caused by residual phase offsets $\phi_m$. Specifically, we assume $\{\phi_m: m=1,2,...,M\}$ are uniformly distributed in $(0,\phi)$ and $\phi$ is the maximum phase offset. That is, $\phi_m\sim U(0,\phi)$.
\item Without loss of generality, the symbol duration is set to $T = 1$. Recall that the p-ML and p-LMMSE estimators make use of only the outputs of the $M$-th matched filter. For these two estimators, the estimation performance hinges on the length of the $M$-th matched filter $d_M=1-\tau_M$. In view of this, the time offsets $\tau_m$, $\forall m$, are set in the following manner: first, we fix the time offset of the $M$-th device $\tau_M$ (and hence $d_M$); then, we generate the time offsets of other devices uniformly in $(0,\tau_M)$.
\item EsN0 is defined as
\begin{eqnarray}
\text{EsN0}\triangleq\frac{1}{N_0}\frac{1}{L}\sum_{i=1}^{L}\left|\sum_{m=1}^{M}e^{j\phi_m}s_m[i] \right|^2.
\end{eqnarray}
\end{enumerate}

The transmitted symbols of the four devices $\bm{s_m}$ are generated uniformly in ranges $[-6,0]$, $[-4,2]$, $[-2,4]$, $[0,6]$, respectively.
Considering the intensive computational complexity of the LMMSE estimator (see \eqref{eq:LMMSE_misaligned}), we use a short packet length $L=128$ to simulate its performance. For all other estimators, the packet length is set to $L=1024$.

\subsection{Synchronous OAC}
We first consider the synchronous OAC and compare the ML and LMMSE estimators under various degrees of phase misalignments and EsN0. The maximum phase offset $\phi$ is set to $0$ (no phase misalignment), $\pi/2$ (mild),\footnote{Notice that $\phi$ is the maximum phase offset and the phase offsets of all devices are uniformly distributed in $[0,\phi]$. If we look at the phase misalignment between any two devices, the average pair\-wise-pha\-se-mis\-alignm\-ent is only $\phi/3$. That is why we classify $\pi/2$ as mild because the average pa\-irwi\-se-ph\-ase-m\-isali\-gn\-ment is only $\pi/6$.} $\pi$ (moderate), or $2\pi$ (severe).

\begin{figure}[t]
  \centering
  \includegraphics[width=0.8\columnwidth]{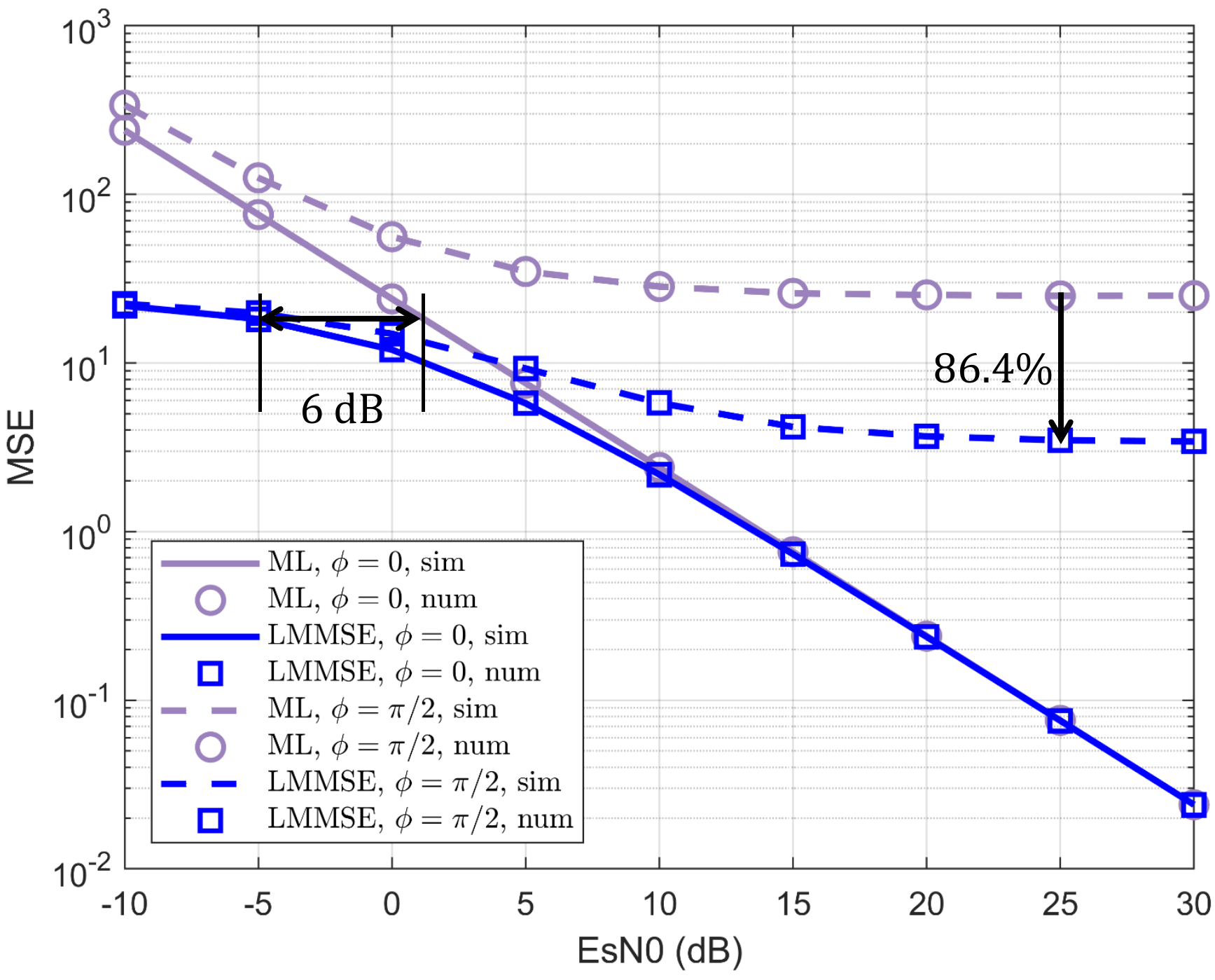}\\
  \caption{Numerical and simulation results of the ML and LMMSE estimators in synchronous OAC under various degrees of phase misalignments.}
\label{fig:sim1}
\end{figure}

Fig.~\ref{fig:sim1} presents the MSEs of the ML and LMMSE estimators versus EsN0 (in dB), wherein $\phi=0$ and $\pi/2$. The numerical MSEs are generated by \eqref{eq:MSENaiveh} and \eqref{eq:MSELMMSEh}, respectively. In all our simulations, the numerical results match the simulation results very well. To ease presentation, we shall omit the numerical results and present the simulation results in the following.

Two main observations from Fig.~\ref{fig:sim1} are as follows:
\begin{enumerate}
\item In the aligned OAC ($\phi=0$), our LMMSE estimator outperforms the ML estimator by much in the low-EsN0 regime. At an EsN0 of $-5$ dB, the MSE gains are up to $6$ dB. In the high-EsN0 regime, the two estimators are equally optimal. This is consistent with our analysis in Section \ref{sec:III}.
\item When there is phase misalignment, both the ML and LMMSE estimators suffer from error floors in the high-EsN0 regime. In particular, the error floor of the aligned-sample estimator is fairly pronounced even with mild phase offset $\phi=\pi/2$. The LMMSE estimator, on the other hand, lowers the error floor by $86.4\%$.
\end{enumerate}

If we further increase $\phi$ to $\pi$ and $2\pi$, similar results can be observed and the LMMSE estimator consistently outperforms the ML estimator. Overall, we conclude that the prior information is very helpful in the synchronous OAC, the MSE performance is improved remarkably with the LMMSE estimator.

\subsection{Asynchronous OAC}
Next, we evaluate the MSEs of the ML and Bayesian estimators designed for the asynchronous OAC.
With no prior information, the viable estimators are the p-ML estimator (Corollary~\ref{thm:Asyn_naiveLMMSE}) and the ML estimator (Definition~\ref{defi:2}).
With prior information, this paper devised a p-LMMSE estimator (Corollary~\ref{thm:Asyn_naiveLMMSE}), an LMMSE estimator (Theorem~\ref{thm:prop7}), and an SP-MAP estimator (Definition~\ref{defi:SP_MAP}).

\subsubsection{The p-ML and the p-LMMSE estimators}
\begin{figure}[t]
  \centering
  \includegraphics[width=0.73\columnwidth]{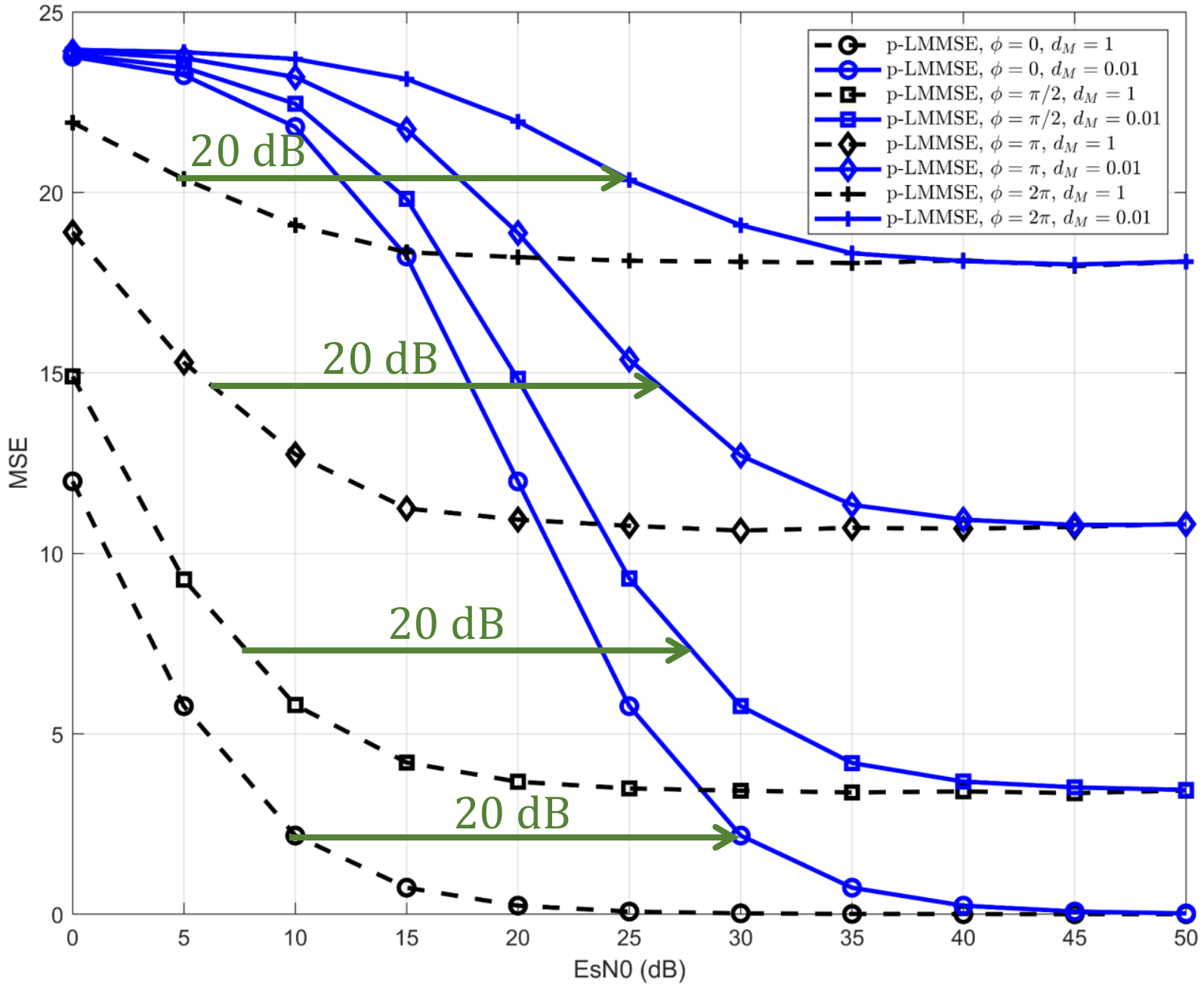}\\
  \caption{MSEs of the p-LMMSE estimator in asynchronous OAC under various degrees of time and phase misalignments.}
\label{fig:sim2}
\end{figure}
The p-ML and p-LMMSE estimators utilize only the outputs of the $M$-th matched filter. Thus, compared with their performance in the synchronous OAC (i.e., Fig.~\ref{fig:sim1}), the introduction of time offset simply results in an EsN0 penalty. For example, if the maximum time offset $\tau_M=0.9$ (hence $d_M=1-\tau_M=0.1$), then we only need to shift the curves of the ML/LMMSE estimators in Fig.~\ref{fig:sim1} by $10$ dB to the right, where $10$ dB is calculated from $10 \log_{10}(1/d_M)$. An immediate result is that the p-LMMSE estimator is still strictly better than the p-ML estimator after the right shift.

Fig.~\ref{fig:sim2} presents the MSE performance of the p-LMMSE estimator (the performance of the p-ML estimator is omitted). As predicted, the MSE performance deteriorates when there is either time or phase misalignment -- time misalignment introduces a $20$ dB EsN0 penalty ($d_M = 0.01$ corresponds to $20$ dB) while phase misalignment results in both EsN0 penalty and error floor.

\subsubsection{The ML estimator}
\begin{figure}[t]
  \centering
  \includegraphics[width=0.79\columnwidth]{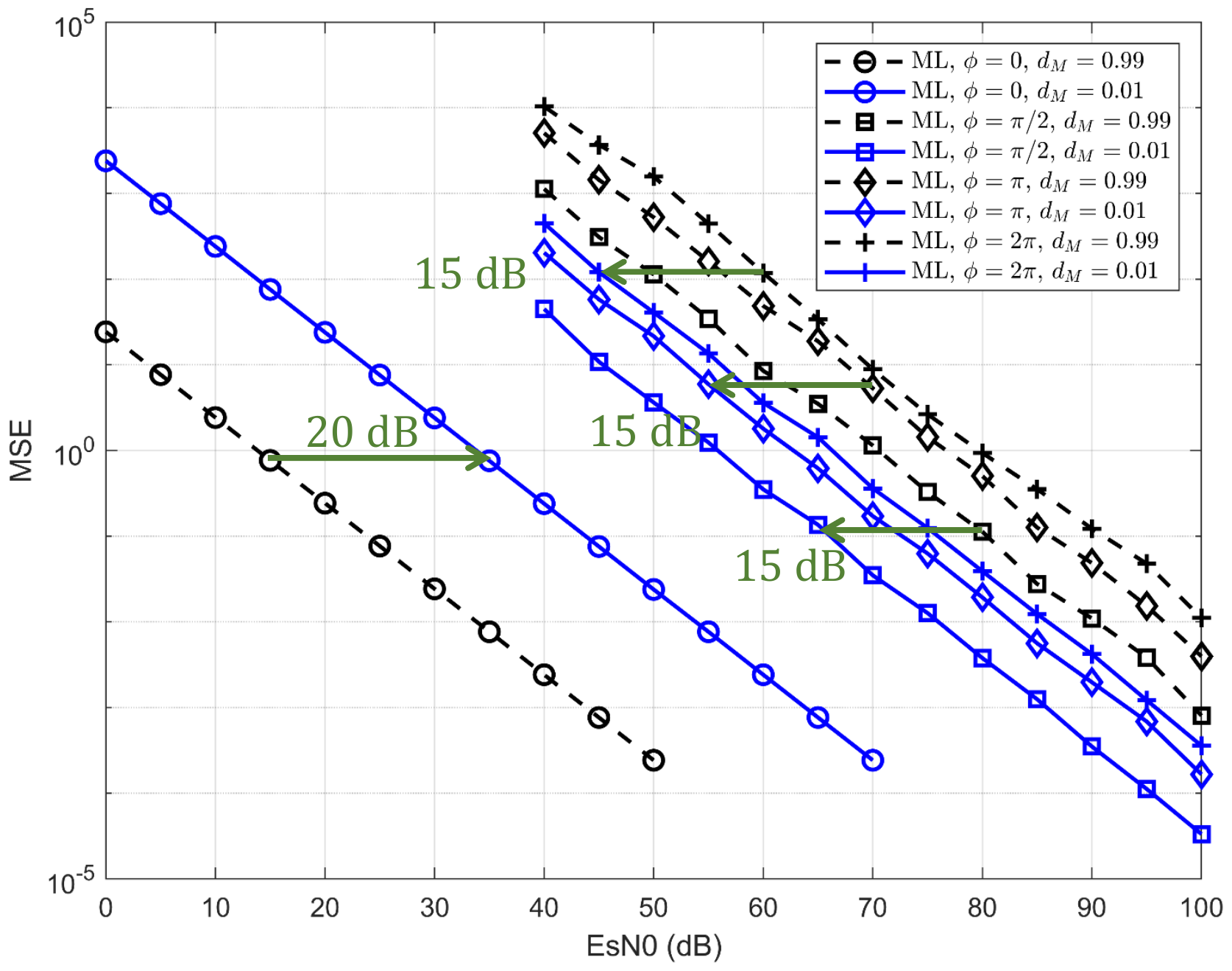}\\
  \caption{MSEs of the ML estimator in asynchronous OAC under various degrees of time and phase misalignments.}
\label{fig:sim3}
\end{figure}
Fig.~\ref{fig:sim3} presents the MSEs of the ML estimator under different time and phase misalignments.
As can be seen, when there is no phase misalignment, the ML estimator suffers from time offset, and the MSE performance deteriorates by $20$ dB when we decrease $d_M$ from $0.99$ to $0.01$. When there is phase misalignment, on the other hand, an interesting observation is that the ML estimator benefits from time misalignment: the MSEs are improved by $15$ dB when $\phi = \pi/2$, $\pi$, and $2\pi$.

ML uses the samples from all the matched filters, but it utilizes no prior information. Comparing Fig.~\ref{fig:sim2} with Fig.~\ref{fig:sim3}, it can be seen that the ML estimator is even worse than the p-LMMSE estimator (which uses only aligned samples) when there is phase misalignment. For example, to achieve a MSE of $10$ when $\phi=\pi/2$ and $\tau_M=0.01$, the p-LMMSE estimator requires an EsN0 of $24$ dB while the ML estimator requires $45$ dB. On the other hand, the advantage of the ML estimator is that it exhibits no error floor in the high-EsN0 regime. The reason is that, utilizing the misaligned samples provides more equations to disentangle the source symbols $\bm{s}$, and hence, the error floor is eliminated.

To summarize, a major problem of the ML estimator is that it is very sensitive to noise due to the infinite estimation space. As a result, it suffers from severe error propagation and noise enhancement when there is phase misalignment \cite{techFL}. This is also validated in Fig.~\ref{fig:sim3}: there is a large EsN0 gap between the phase-aligned OAC and the phase-misaligned OAC.

Next, we evaluate the MSEs of our LMMSE and SP-MAP estimators designed for the asynchronous OAC.

\subsubsection{The SP-MAP and LMMSE estimators}
\begin{figure}[t]
  \centering
  \includegraphics[width=0.76\columnwidth]{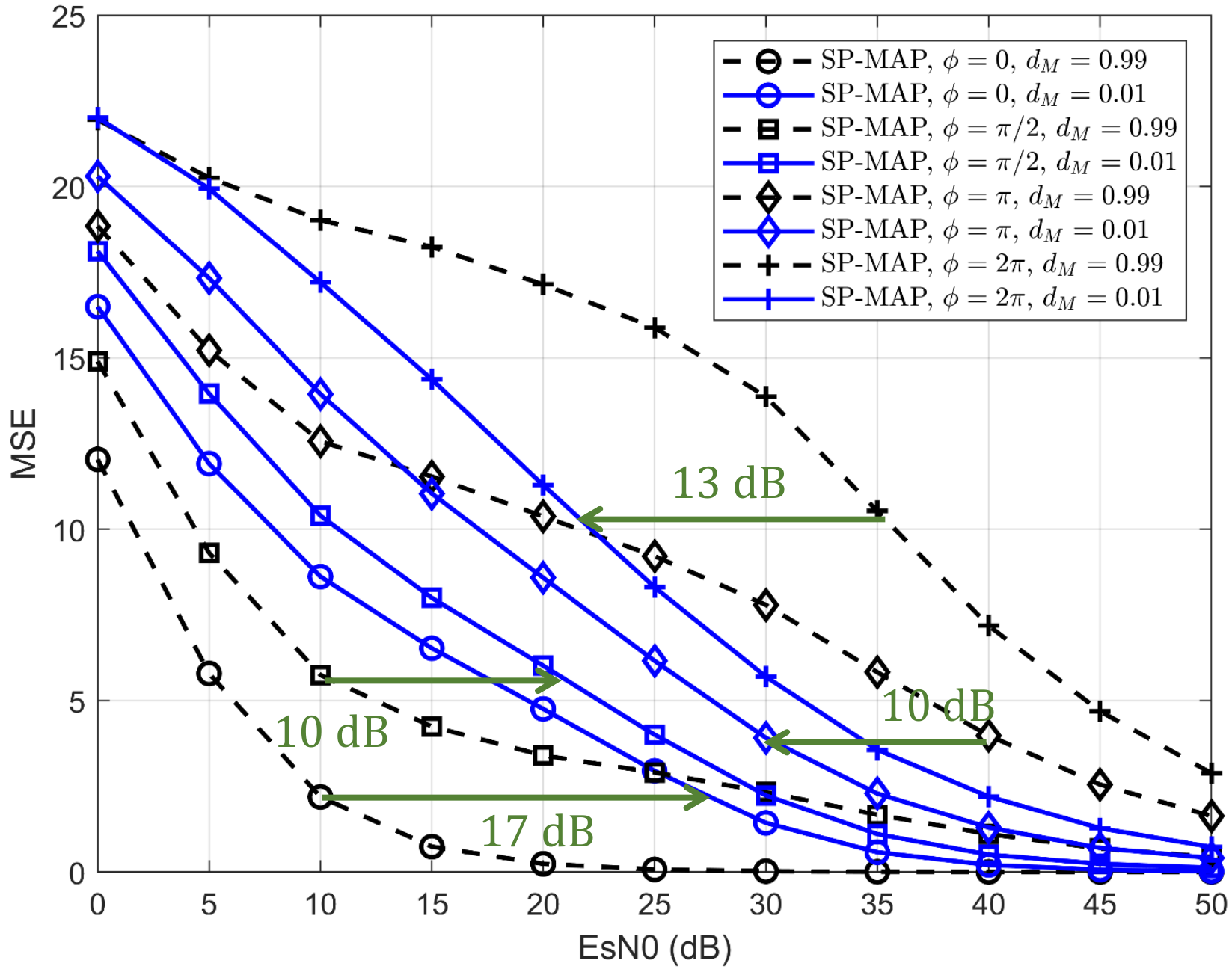}\\
  \caption{MSEs of the SP-MAP estimator in asynchronous OAC under various degrees of time and phase misalignments.}
\label{fig:sim4}
\end{figure}

\begin{figure}[t]
  \centering
  \includegraphics[width=0.76\columnwidth]{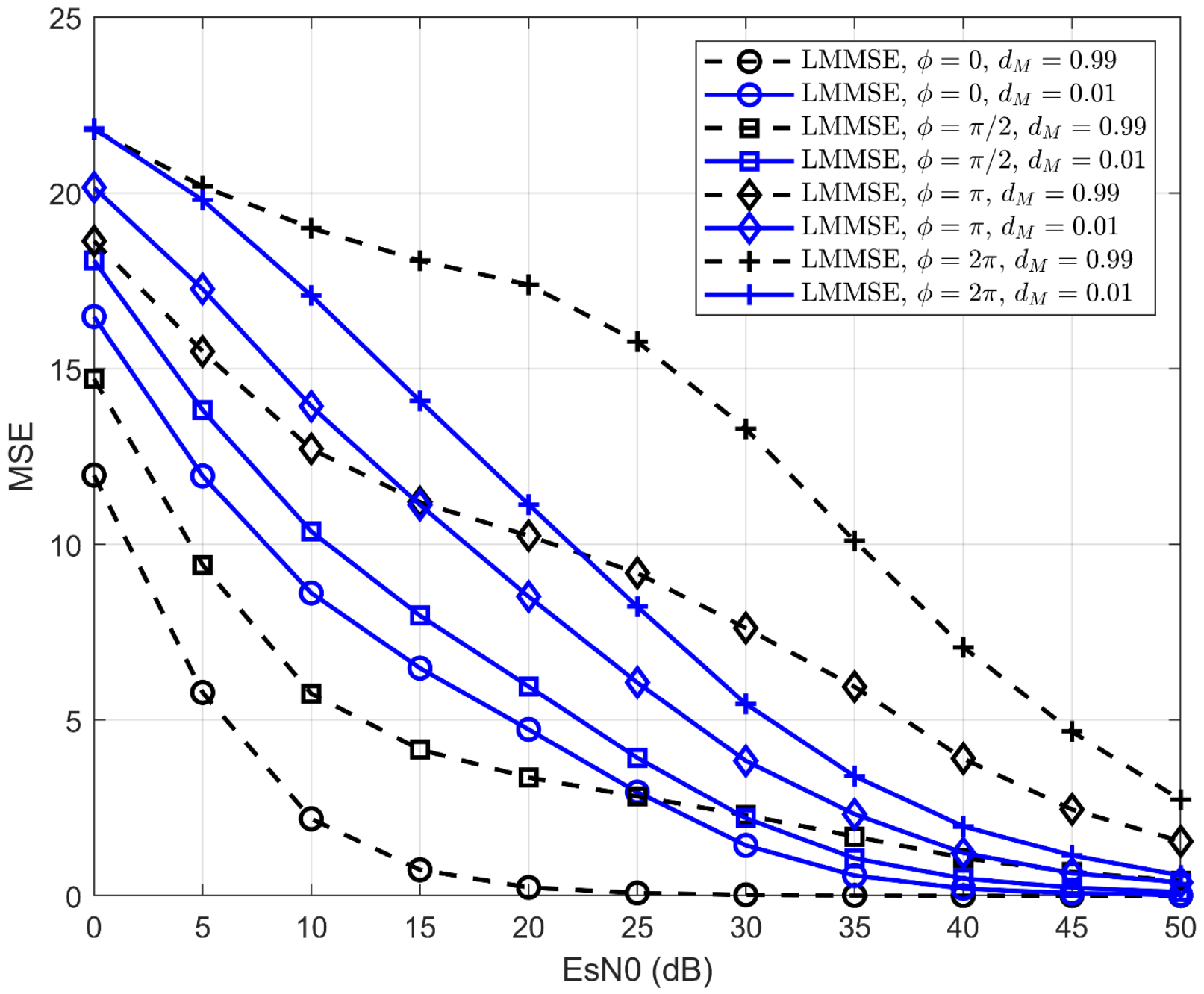}\\
  \caption{MSEs of the LMMSE estimator in asynchronous OAC under various degrees of time and phase misalignments.}
\label{fig:sim5}
\end{figure}
The SP-MAP and LMMSE estimators utilize all the samples, and also, the prior information transmitted from the edge devices. Their MSEs are presented in Fig.~\ref{fig:sim4} and \ref{fig:sim5}, respectively, under various degrees of time and phase misalignments.
As shown, the estimation performance of the two estimators is on the same footing in terms of MSE. Thus, we shall focus on the SP-MAP estimator in the following.

We have two observations from Fig.~\ref{fig:sim4}:
\begin{enumerate}
\item When there is no phase misalignment ($\phi=0$), the EsN0 penalty caused by asynchrony is $17$ dB. Compared with Fig.~\ref{fig:sim2} and Fig.~\ref{fig:sim3}, where the EsN0 penalty is $20$ dB with the p-LMMSE and ML estimators, the SP-MAP estimator compensates the EsN0 penalty introduced by asynchrony by $3$ dB.
\item When there are phase misalignments, the SP-MAP estimator also benefits from the time misalignment. Take the $\phi=\pi$ and $2\pi$ curves in Fig.~\ref{fig:sim5} for example. When there is time misalignment, the MSEs of the SP-MAP estimator are improved by $10$ dB and $13$ dB, respectively.
\end{enumerate}

In summary, 1) compared with the p-LMMSE estimator, the SP-MAP estimator utilizes all the sampling output of the matched filters and completely eliminates the error floors faced by the p-LMMSE estimator;
2) compared with the ML estimator, the SP-MAP estimator addresses the error propagation problem by taking advantage of the prior information transmitted from the devices. The MSEs are significantly reduced in all cases;
3) compared with the LMMSE estimator, the SP-MAP estimator attains the same level of MSE performance, but is much more computationally efficient.

\section{Conclusion}\label{sec:Conclusion}
OAC is an efficient scheme to speed up the function computation in multi-access edge computing.
The main spirit of OAC is joint comput\-ation-and-commun\-ication by exploiting the superposition property of the MAC whose output is an arithmetic sum of the input signals.
Due to the non-viability of accurate channel-gain compensation and perfect synchronization, the OAC system of practical interest is the misaligned OAC with channel-gain mismatches and time asynchronies among edge devices.
An urgent and challenging problem is how to accurately and efficiently estimate the arithmetic sum from the misaligned signals.

This paper put forth a Bayesian approach to solve the estimation problem in the misaligned OAC. A main ingredient of the estimators devised in this paper is the statistical information of the distributed data.
In digital communications, the transmitted symbols are discrete constellations and the detection space is naturally narrowed down to the possible constellation points -- the discrete constellation itself serves as a kind of prior information to the receiver.
In OAC, however, the transmitted symbols are continuous complex values and the estimation space is infinitely large.
In this context, additional prior information can be conducive to narrowing down the estimation space to enable much better estimation performance, especially for the misaligned OAC that suffers from error propagation and noise enhancement.
Among the proposed estimators, our SP-MAP estimator was demonstrated to be the most promising in terms of both MSE performance and computational complexity, paving the way to integrate OAC into the overall picture of multi-tier computing for future wireless networks.



\appendices
\section{}\label{sec:AppA}
This appendix proves that the estimator in \eqref{eq:LMMSEEstimator} is an LMMSE estimator, and derives its MSE.

Given the samples in \eqref{eq:III1}, an LMMSE estimator estimates the sequence $\bm{s}_+\in\mathcal{C}^L$ symbol-by-symbol by
$\widehat{s}_+[i] = \lambda r[i]+c$,
where the constants $\lambda,c\in\mathbb{C}$ are chosen so that the MSE $\frac{1}{L}\sum_{i=1}^L \left|\widehat{s}_+[i]-s_+[i]\right|^2$ is minimized.

First, the MSE of a linear estimator is given by
\small
\begin{eqnarray}\label{eq:III2}
\hspace{-0.5cm}&& \text{MSE}=\frac{1}{L}\sum_{i=1}^L \left|\sum_{m=1}^M (\lambda h_m-1)s_m[i]+c+\lambda z[i]\right|^2 \nonumber\\
\hspace{-0.5cm}&& = \frac{1}{L}\sum_{i=1}^L \left|\sum_{m=1}^M (\lambda h_m\!-\!1)s_m[i]\right|^2 \!\!+ c\frac{1}{L}\sum_{i=1}^L\left[\sum_{m=1}^M (\lambda h_m\!-\!1)s_m[i]\right]^* \nonumber\\
\hspace{-0.5cm}&& \qquad +~ c^*\frac{1}{L}\sum_{i=1}^L\left[\sum_{m=1}^M (\lambda h_m\!-\!1)s_m[i]\right] + |\lambda|^2\frac{N_0}{T} + |c|^2 \nonumber\\
\hspace{-0.5cm}&& = (\lambda\bm{h\!-\!1})^H\frac{1}{L}\sum_{i=1}^L\!\bm{s^*[i]}\bm{s[i]}^\top\!\!(\lambda\bm{h\!-\!1})\!+\!c(\lambda\bm{h\!-\!1})^H\frac{1}{L}\!\sum_{i=1}^L\!\bm{s^*[i]} \nonumber\\
\hspace{-0.5cm}&& \qquad +~ c^*(\lambda\bm{h-1})^\top\frac{1}{L}\sum_{i=1}^L\bm{s[i]} + |\lambda|^2\frac{N_0}{T} + |c|^2 \nonumber\\
\hspace{-0.5cm}&& = (\lambda\bm{h\!-\!1})^H \bm{V}(\lambda\bm{h\!-\!1})\!+\!c(\lambda\bm{h\!-\!1})^H \bm{\widehat{\mu}}^* + c^*(\lambda\bm{h-1})^\top\bm{\widehat{\mu}} \nonumber\\
\hspace{-0.5cm}&& \qquad +~ |\lambda|^2\frac{N_0}{T} + |c|^2,
\end{eqnarray}
\normalsize
where $\bm{V}$ is as defined in \eqref{eq:V} and $\bm{\widehat{\mu}}$ as in \eqref{eq:mu}. As can be seen, MSE is a quadratic function of both $\lambda$ and $c$. The optimal $\lambda$ and $c$ that minimize the MSE can be obtained by setting:
\begin{eqnarray}
\label{eq:partial1}
\hspace{-1cm}&& \frac{\partial \text{MSE}}{\partial \lambda} = \lambda^*\bm{h}^H\bm{Vh} - \bm{1}^\top\bm{Vh} + c^*\bm{h}^\top\bm{\widehat{\mu}} + \lambda^*\frac{N_0}{T}=0, \\
\label{eq:partial2}
\hspace{-1cm}&& \frac{\partial \text{MSE}}{\partial c} = (\lambda\bm{h-1})^H \bm{\widehat{\mu}}^* + c^* = 0.
\end{eqnarray}

Substituting \eqref{eq:partial2} into \eqref{eq:partial1} yields
\begin{eqnarray*}
\lambda = \frac{\bm{h}^H\left(\bm{V-\bm{\widehat{\mu}}^*\bm{\widehat{\mu}}^\top}\right)\bm{1}}{\bm{h}^H\left(\bm{V-\bm{\widehat{\mu}}^*\bm{\widehat{\mu}}^\top}\right)\bm{h}+\frac{N_0}{T}}.
\end{eqnarray*}

To simplify $\lambda$, let us further define $\bm{D}=\bm{V-\bm{\widehat{\mu}}^*\bm{\widehat{\mu}}^\top}$
and we finally have
\begin{eqnarray*}
\lambda = \frac{\bm{h}^H\bm{D}\bm{1}}{\bm{h}^H\bm{D}\bm{h}+\frac{N_0}{T}},~~c=\left(\bm{1}\!-\!\frac{\bm{h}^H\bm{D1}}{\bm{h}^H\bm{Dh}+\frac{N_0}{T}}\bm{h}\right)^\top \!\!\bm{\widehat{\mu}}.
\end{eqnarray*}
Thus, \eqref{eq:LMMSEEstimator} is the LMMSE estimator that minimizes the MSE. Notice that this is an unbiased estimator since
\begin{equation*}
\frac{1}{L}\sum_{i=1}^L(\widehat{s}_+[i]\!-\!{s}_+[i])= \lambda\bm{h}^\top\bm{\widehat{\mu}} \!+\! (\bm{1}\!-\!\lambda\bm{h})^\top\bm{\widehat{\mu}}\!-\!\bm{1}^\top\bm{\widehat{\mu}}=0.
\end{equation*}

Substituting $\lambda$ and $c$ back into \eqref{eq:III2} yields
\begin{eqnarray*}
\text{MSE} &&\hspace{-0.5cm}= (\lambda\bm{h\!-\!1})^H \bm{V}(\lambda\bm{h\!-\!1})\!-\!(\lambda\bm{h\!-\!1})^\top\bm{\widehat{\mu}}\bm{\widehat{\mu}}^H (\lambda\bm{h\!-\!1})^*   \\
&&\hspace{-1.5cm} \qquad - (\lambda\bm{h\!-\!1})^H\bm{\widehat{\mu}}^*\bm{\widehat{\mu}}^\top (\lambda\bm{h\!-\!1}) + |\lambda|^2\frac{N_0}{T} + |(\lambda\bm{h\!-\!1})^\top\bm{\widehat{\mu}}|^2 \\
&&\hspace{-0.5cm}= (\lambda\bm{h\!-\!1})^H \bm{D}(\lambda\bm{h\!-\!1}) + |\lambda|^2\frac{N_0}{T} \\
&&\hspace{-0.5cm}= \bm{1}^\top\bm{D1}-\frac{\left|\bm{h}^H\bm{D1}\right|^2}{\bm{h}^H\bm{Dh}+\frac{N_0}{T}}.
\end{eqnarray*}

We next compare $\text{MSE}_\text{LMMSE}$ with $\text{MSE}_\text{ML}$. From \eqref{eq:MSENaiveh} and \eqref{eq:MSELMMSEh}, we have
\begin{eqnarray}\label{eq:III3}
\hspace{-0.5cm}&&\text{MSE}_\text{ML} - \text{MSE}_\text{LMMSE} \\
\hspace{-0.5cm}&&= (\bm{h\!-\!1})^H\bm{V}(\bm{h\!-\!1})\!+\!\frac{N_0}{T} \!-\! \left(\bm{1}^\top\bm{D1}\!-\!\frac{\left|\bm{h}^H\bm{D1}\right|^2}{\bm{h}^H\bm{Dh}\!+\!\frac{N_0}{T}} \right). \nonumber
\end{eqnarray}

Multiplying both sides of \eqref{eq:III3} by $\bm{h}^H\bm{Dh}+\frac{N_0}{T}$ and defining
\begin{eqnarray}\label{eq:III4}
q\left(\frac{N_0}{T}\right) \triangleq&&\hspace{-0.55cm} \left(\text{MSE}_\text{ML} - \text{MSE}_\text{LMMSE} \right) \left(\bm{h}^H\bm{Vh}+\frac{N_0}{T} \right) \nonumber\\
=&&\hspace{-0.55cm} \Big(\bm{h}^H\bm{Dh}+\frac{N_0}{T} \Big) \Big(\bm{h}^H\bm{Vh}-\bm{1}^\top\bm{Vh}-\bm{h}^H\bm{V1} \nonumber\\
&&\hspace{-0.5cm} +\bm{1}^\top\bm{V1}-\bm{1}^\top\bm{D1}+\frac{N_0}{T} \Big) + |\bm{1}^\top\bm{Dh}|^2.
\end{eqnarray}
Since $\bm{D}$ is positive definite, we have $\bm{h}^H\bm{Dh}+\frac{N_0}{T}>0$. To prove $\text{MSE}_\text{LMMSE}\leq \text{MSE}_\text{ML}$, we only need to prove the minimum value of $q\left(\frac{N_0}{T}\right)$ is nonnegative.

From \eqref{eq:III4}, we know that $q(N_0/T)$ is a quadratic function of $N_0/T$. Then, $N_0/T$ that minimizes $q(N_0/T)$ can be obtained by setting:
\begin{eqnarray*}
\hspace{-0.5cm}&& \frac{\partial q(N_0/T)}{\partial (N_0/T)} = 2\frac{N_0}{T} + \bm{h}^H\bm{Dh} + \bm{h}^H\bm{Vh} - \bm{1}^\top\bm{Vh} - \bm{h}^H\bm{V1} \\
\hspace{-0.5cm}&& \hspace{2cm}  +~ \bm{1}^\top\bm{V1} - \bm{1}^\top\bm{D1} =0.
\end{eqnarray*}

Since the noise variance cannot be negative, we have
\begin{eqnarray*}
\frac{N_0}{T} =&&\hspace{-0.55cm} \max\Big(0,\frac{1}{2}\big[\bm{1}^\top\bm{D1} - (\bm{h}^H\bm{Dh} + \bm{h}^H\bm{Vh} - \bm{1}^\top\bm{Vh} \\
&&\hspace{-0.55cm} - \bm{h}^H\bm{V1} + \bm{1}^\top\bm{V1}) \big]\Big).
\end{eqnarray*}

1) When $\frac{1}{2}\big[\bm{1}^\top\bm{D1} - (\bm{h}^H\bm{Dh} + \bm{h}^H\bm{Vh} - \bm{1}^\top\bm{Vh} - \bm{h}^H\bm{V1} + \bm{1}^\top\bm{V1}) \big] \leq 0$, we have $N_0/T = 0$ and
\begin{eqnarray*}
\hspace{-0.65cm}&& q\left(\frac{N_0}{T}\right)\geq q(0)= \bm{h}^H\bm{Dh}(\bm{h}^H\bm{Vh}-\bm{1}^\top\bm{Vh}-\bm{h}^H\bm{V1} \\
\hspace{-0.65cm}&& \qquad\qquad +~ \bm{1}^\top\bm{V1}) - \bm{h}^H\bm{Dh}\bm{1}^\top\bm{D1}+|\bm{1}^\top\bm{Dh}|^2
\end{eqnarray*}

Let us define $\bm{E}\triangleq\bm{V-D}=\bm{\widehat{\mu}}^*\bm{\widehat{\mu}}^\top$, then
\begin{eqnarray*}
\hspace{-0.65cm}&& q(0)\!=\! \bm{h}^H\!\bm{Dh}(\bm{h}^H\bm{Vh}\!-\!\bm{1}^\top\!\bm{Vh}\!-\!\bm{h}^H\!\bm{V1}\!+\!\bm{1}^\top\!\bm{E1})\!+\!|\bm{1}^\top\!\bm{Dh}|^2 \\
\hspace{-0.65cm}&& = \bm{h}^H\bm{Dh}(\bm{h}^H\bm{Dh} \!-\! \bm{1}^\top\bm{Dh} \!-\! \bm{h}^H\bm{D1}) + \bm{h}^H\bm{Dh}(\bm{h}^H\bm{Eh} \\
\hspace{-0.65cm}&&  \qquad -~ \bm{1}^\top\bm{Eh} \!-\! \bm{h}^H\bm{E1} \!+\! \bm{1}^\top\bm{E1})+|\bm{1}^\top\bm{Dh}|^2 \\
\hspace{-0.65cm}&& = (\bm{h\!-\!1})^H\bm{Dh}\bm{h}^H\bm{D}(\bm{h\!-\!1}) \!+\! \bm{h}^H\bm{Dh}(\bm{h\!-\!1})^H\bm{E}(\bm{h\!-\!1}) \\
\hspace{-0.65cm}&& = |(\bm{h\!-\!1})^H\bm{Dh}|^2 +\bm{h}^H\bm{Dh}|(\bm{h\!-\!1})^\top\bm{\widehat{\mu}}|^2 \geq 0,
\end{eqnarray*}
where the last inequality follows because $\bm{D}$ is positive definite.
Therefore,
\begin{eqnarray*}
\text{MSE}_\text{ML}- \text{MSE}_\text{LMMSE}
\geq \frac{q(0)}{\bm{h}^H\bm{Dh}}\geq 0.
\end{eqnarray*}
This formula matches our intuition: when the noise variance $N_0/T=0$ and the channel precoding is perfect, i.e., $\bm{h=1}$, we have $q(0)=0$. The ML and the LMMSE estimators are the same in this case as $\lambda=1$ and $c=0$.


2) When $\frac{1}{2}\big[\bm{1}^\top\bm{D1} - (\bm{h}^H\bm{Dh} + \bm{h}^H\bm{Vh} - \bm{1}^\top\bm{Vh} - \bm{h}^H\bm{V1} + \bm{1}^\top\bm{V1}) \big] > 0$, we have
\begin{eqnarray*}
\hspace{-0.5cm}&& q\left(\frac{N_0}{T}\right)\geq q\Big(\frac{1}{2}\big[\bm{1}^\top\bm{D1} - (\bm{h}^H\bm{Dh} + \bm{h}^H\bm{Vh} - \bm{1}^\top\bm{Vh} \\
\hspace{-0.5cm}&& \qquad\qquad -~ \bm{h}^H\bm{V1} + \bm{1}^\top\bm{V1}) \big]\Big) \\
\hspace{-0.5cm}&& = |\bm{1}^\top\bm{Dh}|^2 -\frac{1}{4}\Big(\bm{1}^\top\bm{Vh}+\bm{h}^H\bm{V1}-\bm{h}^H\bm{Eh}-\bm{1}^\top\bm{E1}  \Big)^2 \\
\hspace{-0.5cm}&& = |\bm{1}^\top\bm{Dh}|^2 -\frac{1}{4}\Big[\bm{1}^\top\bm{Dh}+\bm{h}^H\bm{D1}\!-\!(\bm{h\!-\!1})^H\bm{E}(\bm{h\!-\!1})  \Big]^2 \\
\hspace{-0.5cm}&& = |\bm{1}^\top\bm{Dh}|^2 - \frac{1}{4}\Big[ 2(\bm{1}^\top\bm{Dh})^\mathfrak{r} - |(\bm{h\!-\!1})^\top\bm{\widehat{\mu}}|^2 \Big]^2 \\
\hspace{-0.5cm}&& \overset{(a)}{\geq} |\bm{1}^\top\bm{Dh}|^2 -\left[(\bm{1}^\top\bm{Dh})^\mathfrak{r} \right]^2= \left[(\bm{1}^\top\bm{Dh})^\mathfrak{i} \right]^2 \geq 0
\end{eqnarray*}
where (a) follows from $\frac{1}{2}\big[\bm{1}^\top\bm{D1} - (\bm{h}^H\bm{Dh} + \bm{h}^H\bm{Vh} - \bm{1}^\top\bm{Vh} - \bm{h}^H\bm{V1} + \bm{1}^\top\bm{V1}) \big] > 0$. That is, we have $\bm{1}^\top\bm{Vh}+\bm{h}^H\bm{V1}-\bm{h}^H\bm{Eh}-\bm{1}^\top\bm{E1}\geq 0$, and hence $2(\bm{1}^\top\bm{Dh})^\mathfrak{r} - |(\bm{h\!-\!1})^\top\bm{\widehat{\mu}}|^2>0$.

Overall, we have $\text{MSE}_\text{LMMSE}\leq \text{MSE}_\text{ML}$.

\section{}\label{sec:AppA2}
This appendix proves Theorem \ref{thm:prop7}. 
We first show that the estimator in \eqref{eq:LMMSE_misaligned} is an LMMSE estimator.
Given the signal model in \eqref{eq:samplesMat}, a linear estimator estimates $\bm{s}_+$ by
$\widehat{\bm{s}}_+=\bm{A y+c}$.
The MSE of the linear estimate $\widehat{\bm{s}}_+$ is then given by
\begin{eqnarray}\label{eq:MSE_mislaigned}
\text{MSE}=\frac{1}{L}\mathbb{E}\left[(\bm{Ay+c}-{\bm{s}}_+)^H(\bm{Ay+c}-{\bm{s}}_+)\right].
\end{eqnarray}

The matrix $\bm{A}$ and vector $\bm{c}$ that yield the minimum MSE can then be obtained by setting $\partial\text{MSE}/\partial\bm{A}=0$ and $\partial\text{MSE}/\partial\bm{c}=0$. Thus, we have
\begin{eqnarray*}
\hspace{-0.65cm}&&\frac{\partial\text{MSE}}{\partial\bm{A}}=\mathbb{E}\frac{\partial\text{Tr}[(\bm{Ay+c}-{\bm{s}}_+)(\bm{Ay+c}-{\bm{s}}_+)^H]}{\partial\bm{A}} \\
\hspace{-0.65cm}&&=\!\bm{A}\mathbb{E}[\bm{yy^H}] +\bm{c}\mathbb{E}^H[\bm{y}]-\mathbb{E}[\bm{s_+}\bm{y^H}]\\
\hspace{-0.65cm}&&=\!\! \bm{AG\mathbb{E}[ss^H]G^H\!\!+\!\!A\Sigma_z}\!+\!\bm{c\mathbb{E}^H[s]G^H}\!\!-\!\bm{F\mathbb{E}[ss^H]G^H}\!\!=\!0, \\
\hspace{-0.65cm}&&\frac{\partial\text{MSE}}{\partial\bm{c}}\!=\!\bm{A\mathbb{E}[y]\!+\!c\!-\!\mathbb{E}[s_+]} \!=\! \bm{AG\mathbb{E}[s]}\!+\!\bm{c}\!-\!\bm{F\mathbb{E}[s]}\!=\!0.
\end{eqnarray*}
Given $\widetilde{\bm{\mu}}=\mathbb{E}[\bm{s}]$ and $\bm{\widetilde{D}}=\mathbb{E}[\bm{ss^H}]-\mathbb{E}[\bm{s}]\mathbb{E}^H[\bm{s}]$, we have
\begin{eqnarray*}
\bm{A}
\hspace{-0.2cm}&=&\hspace{-0.2cm} \bm{F\widetilde{D}G^H}(\bm{G\widetilde{D}G^H}+\bm{\Sigma_z})^{-1}, \\
\bm{c}
\hspace{-0.2cm}&=&\hspace{-0.2cm} \bm{F}\widetilde{\bm{\mu}}-\bm{AG}\widetilde{\bm{\mu}}.
\end{eqnarray*}
This gives us the LMMSE estimator in \eqref{eq:LMMSE_misaligned}. The MSE of the LMMSE estimator can be obtained by substituting $\bm{A}$ and $\bm{c}$ into \eqref{eq:MSE_mislaigned}, giving
\begin{eqnarray*}
&&\hspace{-0.5cm} \text{MSE}=\frac{1}{L}\mathbb{E}\left[(\bm{Ay+c}-{\bm{s}}_+)^H(\bm{Ay+c}-{\bm{s}}_+)\right] \\
&&\hspace{-0.5cm} = \frac{1}{L}\mathbb{E}\left\{\left[(\bm{AG\!-\!F})\bm{s\!+\!Az\!+\!c}\right]^H\left[(\bm{AG\!-\!F})\bm{s\!+\!Az\!+\!c}\right]\right\} \\
&&\hspace{-0.5cm} = \frac{1}{L}\mathbb{E}\big[\bm{s}^H(\bm{AG-F})^H(\bm{AG-F})\bm{s}+\bm{c}^H(\bm{AG-F})\bm{s}+\\
&&\hspace{-0.5cm} \qquad\bm{z}^H\bm{A}^H\bm{Az} + \bm{s}^H(\bm{AG-F})^H\bm{c}+\bm{c}^H\bm{c} \big] \\
&&\hspace{-0.5cm} \overset{(a)}{=} \frac{1}{L}\mathbb{E}\big[\bm{s}^H(\bm{AG-F})^H(\bm{AG-F})\bm{s}+\bm{z}^H\bm{A}^H\bm{Az}\\
&&\hspace{-0.5cm}  \qquad- \frac{1}{L}\widetilde{\bm{\mu}}^H(\bm{AG-F})^H(\bm{AG-F})\widetilde{\bm{\mu}} \big]\\
&&\hspace{-0.5cm} =\! \frac{1}{L}\text{Tr}\left\{ (\bm{AG\!\!-\!\!F})(\mathbb{E}[\bm{ss}^H]\!\!-\!\!\widetilde{\bm{\mu}}\widetilde{\bm{\mu}}^H)(\bm{AG\!\!-\!\!F})^H\!+\!\bm{A\Sigma_zA}^H \right\} \\
&&\hspace{-0.5cm}= \frac{1}{L}\text{Tr}\left[(\bm{AG\!-\!F})\widetilde{\bm{D}}(\bm{AG\!-\!F})^H  \!+\! \bm{A\Sigma_zA}^H \right],
\end{eqnarray*}
where (a) follows by substituting $\bm{c}=(\bm{F}-\bm{AG})\widetilde{\bm{\mu}}$.

\section{}\label{sec:AppB}
This appendix proves Theorem \ref{thm:AMP}.
Note that the key of Theorem \ref{thm:AMP} is the first part that the marginal posterior distribution $f(\bm{s[i]} |\allowbreak \bm{y})$ is an $M$-dimensional complex Gaussian distribution. Provided that this argument is right, the rest of Theorem \ref{thm:AMP} holds.
In the following, let us dive deeper into the analog message passing and prove that $f(\bm{s[i]} |\allowbreak \bm{y})$ is an $M$-dimensional complex Gaussian distribution and can be computed by Algorithm \ref{algo:1}.

To begin with, we point out that a multivariate Gaussian distribution can be parameterized by two sets of parameters \cite{gauvain1994maximum,ahrendt2005multivariate,bromiley2003products}: the moment parameter $(\bm{\mu},\bm{\Sigma})$ and the canonical parament $(\bm{\eta},\bm{\Lambda})$. The two sets of parameters can be transformed into one another and they are useful in different circumstances, as detailed below.

{\bf The moment parameters} $(\bm{\mu},\bm{\Sigma})$. For a multivariate real Gaussian random variable $\bm{w}$ of dimension $2M$, its moment parameters are defined as
\begin{eqnarray*}
\bm{\mu} = \mathbb{E}[\bm{w}],~~\bm{\Sigma} = \mathbb{E}\left[(\bm{w-\mu})(\bm{w-\mu})^\top \right].
\end{eqnarray*}
The moment form of the Gaussian distribution is given by
\begin{small}
\begin{eqnarray*}
\mathcal{N}(\bm{w};\bm{\mu},\!\bm{\Sigma})=\frac{1}{(2\pi)^M|\bm{\Sigma}|^{\frac{1}{2}}} \exp \left\{\!-\frac{1}{2}(\bm{w\!-\!\mu})^\top \bm{\Sigma}^\dagger (\bm{w\!-\!\mu})\!\right\}.
\end{eqnarray*}
\end{small}

\begin{lem}[Marginalization of a multivariate Gaussian \cite{ahrendt2005multivariate}]\label{lemma1}
Let $\bm{w}\sim\mathcal{N}(\bm{w};\bm{\mu},\!\bm{\Sigma})$ be a multivariate Gaussian random variable of dimension $2M$ with the moment parameters being $(\bm{\mu},\bm{\Sigma})$. Let us partition $\bm{w}=[\bm{w_1},\bm{w_2}]^\top$ where $\bm{w_1}$, $\bm{w_2}$ are multivariate Gaussians of dimension $\kappa$ and $2M-\kappa$, respectively. The moment parameters can be partitioned accordingly as
\begin{eqnarray*}
\bm{\mu}=\begin{bmatrix}
\bm{\mu_1} \\
\bm{\mu_2}
\end{bmatrix},~~~
\bm{\Sigma}=\begin{bmatrix}
\bm{\Sigma_{11}} & \bm{\Sigma_{12}} \\
\bm{\Sigma_{21}} & \bm{\Sigma_{22}}
\end{bmatrix}.
\end{eqnarray*}
If we marginalize out $\bm{w_2}$ from $\bm{w}$, the marginal $f(\bm{w_1})$ is still a Gaussian distribution, giving
\begin{eqnarray}
f(\bm{w_1})\hspace{-0.2cm}&=&\hspace{-0.2cm}\int_{\bm{w_2}} \mathcal{N}\left(\begin{bmatrix}
\bm{w_1} \\
\bm{w_2}
\end{bmatrix};\begin{bmatrix}
\bm{\mu_1} \\
\bm{\mu_2}
\end{bmatrix},\!\begin{bmatrix}
\bm{\Sigma_{11}} & \bm{\Sigma_{12}} \\
\bm{\Sigma_{21}} & \bm{\Sigma_{22}}
\end{bmatrix}\right) d\bm{w_2} \nonumber\\
\hspace{-0.2cm}&\propto&\hspace{-0.2cm}
\mathcal{N}(\bm{w_1};\bm{\mu_1},\!\bm{\Sigma_{11}}).
\end{eqnarray}
\end{lem}

{\bf The canonical paraments} $(\bm{\eta},\bm{\Lambda})$. For a Gaussian random variable $\bm{w}$ of dimension $2M$, its canonical parameters are defined as
\begin{eqnarray*}
\bm{\eta} = \Sigma^\dagger \bm{\mu},~~~  \bm{\Lambda} = \Sigma^\dagger.
\end{eqnarray*}

The canonical form of the Gaussian distribution is given by
\begin{eqnarray*}
\mathcal{N}(\bm{w};\bm{\eta},\bm{\Lambda})=\exp \left\{-\frac{1}{2}\bm{w}^\top\bm{\Lambda}\bm{w}+\bm{w}^\top\bm{\eta}+\rho   \right\},
\end{eqnarray*}
where $\rho$ is a constant
\begin{eqnarray*}
\rho = -\frac{1}{2}\left(2M\ln 2\pi -\ln |\bm{\Lambda}| + \bm{\eta}^\top \bm{\Lambda}^\dagger \bm{\eta} \right).
\end{eqnarray*}

\begin{lem}[Product of multivariate Gaussians \cite{bromiley2003products}]\label{lemma2}
Let $\{\bm{w_k}: k=1,2,...,K\}$, $\bm{w_k}\sim\mathcal{N}(\bm{w}; \bm{\eta_k}, \bm{\Lambda_k})$ be a set of multivariate real Gaussian random variable of dimension $2M$. Then, the product of them is still a Gaussian with the new canonical parameters being the sum of the canonical parameters of the original $K$ Gaussians.
\begin{eqnarray}
\hspace{-0.5cm}&& \prod_{k=1}^{K}\mathcal{N}(\bm{w_k};\bm{\eta_k},\bm{\Lambda_k}) \propto \mathcal{N}\left(\bm{w};\sum_{k=1}^K\bm{\eta_k},\sum_{k=1}^K\bm{\Lambda_k}\right) \\
\hspace{-0.5cm}&& \propto \exp \left\{-\frac{1}{2}\bm{w}^\top \sum_{k=1}^K \bm{\Lambda_k}\bm{w} +\bm{w}^\top\sum_{k=1}^K\bm{\eta_k} + \sum_{k=1}^K{\rho_k} \right\}. \nonumber
\end{eqnarray}
\end{lem}

Now that the factor graph in Fig.~\ref{fig:5} has a tree structure, we only need to pass the messages from left to right (forward message passing) and then from right to left (backward message passing). Each message needs to be computed only once, after which the exact marginal posterior distribution converges.

{\it \textbf{Forward Message Passing}} -- We first investigate how the messages are passed from left to right in Fig.~\ref{fig:5}. Without loss of generality, we shall focus on message passing from one variable $\bm{W_{k,i}}$ to another variable $\bm{W_{k+1,i}}$ on the right.

Notice that $\bm{W_{k,i}}=\allowbreak \mathcal{V}(y_k[i])=\{\allowbreak s_1[i],...,\allowbreak s_k[i],\allowbreak s_{k+1}[i-1],\allowbreak s_{k+2}[i-1],\allowbreak ...,s_M[i-1]\}$ and $\bm{W_{k+1,i}}=\allowbreak \mathcal{V}(y_{k+1}[i])=\{\allowbreak s_1[i],...,\allowbreak s_k[i],\allowbreak s_{k+1}[i],\allowbreak s_{k+2}[i-1],\allowbreak ...,s_M[i-1]\}$. Thus, the only difference between $\bm{W_{k,i}}$ and $\bm{W_{k+1,i}}$ is the $(k+1)$-th symbol.
We consider each complex random variable $s_k[i]$ as a real random vector with the elements being the real and imaginary parts. Then, each $\bm{W_{k,i}}$ can be viewed as a $2M$-dimensional real random variable. To simplify the notation, we denote the $2M$-dimensional real variates corresponding to $\bm{W_{k,i}}$ and $\bm{W_{k+1,i}}$, respectively, by

\begin{figure}[t]
  \centering
  \includegraphics[width=0.98\columnwidth]{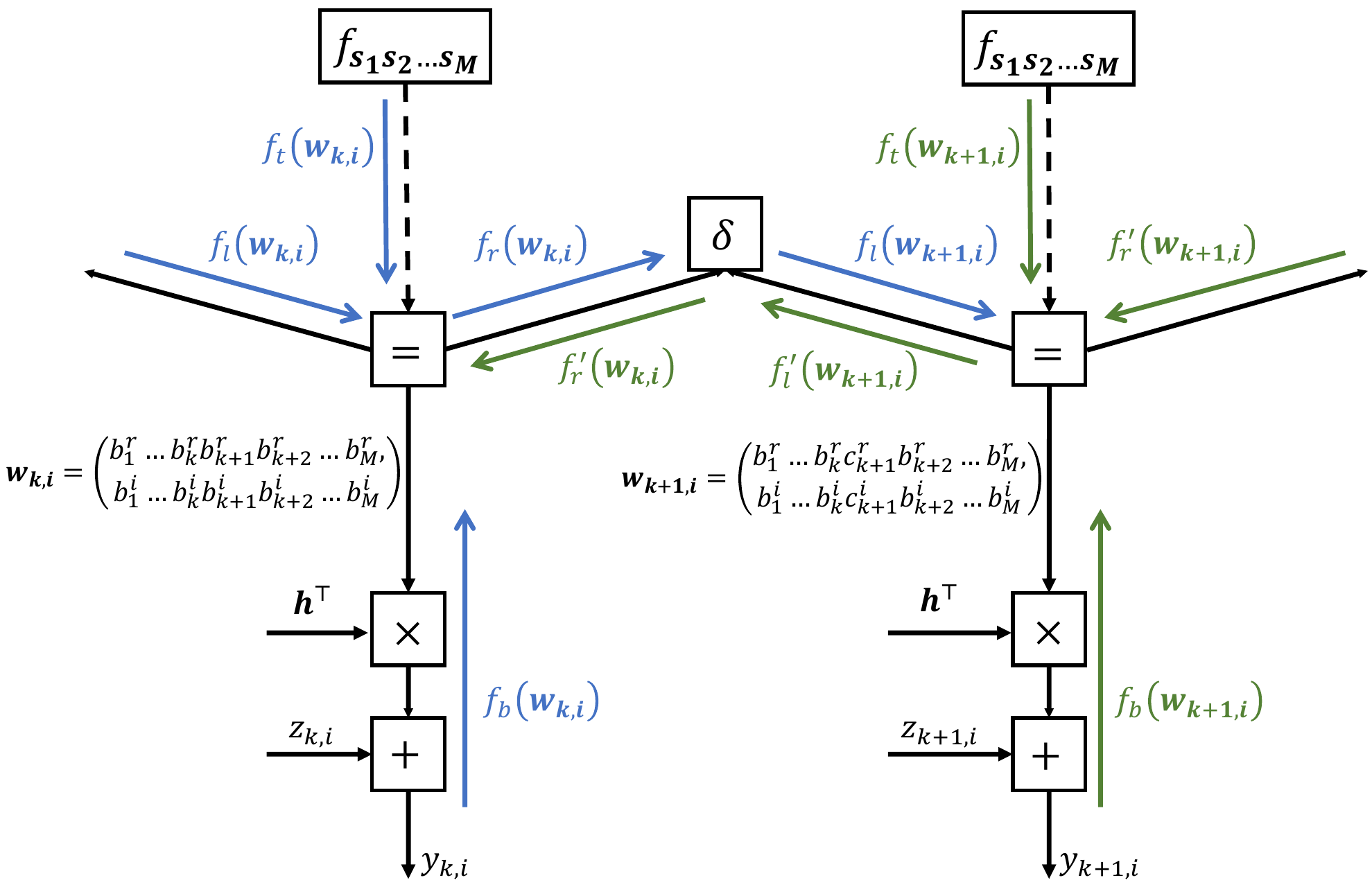}\\
  \caption{The forward message passing from $\bm{W_{k,i}}$ to $\bm{W_{k+1,i}}$ (in blue) and the backward message passing from $\bm{W_{k+1,i}}$ to $\bm{W_{k,i}}$ (in green).}
\label{fig:6}
\end{figure}

\footnotesize
\begin{eqnarray*}
\bm{w_{k,i}} \hspace{-0.3cm}&=&\hspace{-0.3cm}\! \left(\!b^\mathfrak{r}_1,...,b^\mathfrak{r}_k,b^\mathfrak{r}_{k+1},b^\mathfrak{r}_{k+2},...,b^\mathfrak{r}_M,
b^\mathfrak{i}_1,...,b^\mathfrak{i}_k,b^\mathfrak{i}_{k+1},b^\mathfrak{i}_{k+2},...,b^\mathfrak{i}_M \! \right), \\
\bm{w_{k+1,i}} \hspace{-0.3cm}&=&\hspace{-0.3cm}\! \left(\! b^\mathfrak{r}_1,...,b^\mathfrak{r}_k,c^\mathfrak{r}_{k+1},b^\mathfrak{r}_{k+2},...,b^\mathfrak{r}_M,
b^\mathfrak{i}_1,...,b^\mathfrak{i}_k,c^\mathfrak{i}_{k+1},b^\mathfrak{i}_{k+2},...,b^\mathfrak{i}_M \! \right),
\end{eqnarray*}
\normalsize
as shown in Fig.~\ref{fig:6}.

On the left half of Fig.~\ref{fig:6}, there are four edges centered around the equality function ``$=$'' (marked in blue). Due to the equality constraint, these four edges are associated with the same high-dimensional variable $\bm{w_{k,i}}$. In the forward message passing, there are four messages to be computed.

1) The message passed from the bottom, denoted by $f_b(\bm{w_{k,i}})$. This message carries the information about $\bm{w_{k,i}}$ contained in the sample $y_{k,i}$. As per \eqref{eq:samples},
\begin{eqnarray*}
\hspace{-0.65cm} && y_{k,i}=\sum_{m=1}^M (h^\mathfrak{r}_{m}+jh^\mathfrak{i}_{m})(b^\mathfrak{r}_m+jb^\mathfrak{i}_m) + ({z}^\mathfrak{r}_{k,i}  + j{z}^\mathfrak{i}_{k,i}) \\
\hspace{-0.65cm} && =\!\! \sum_{m=1}^M (h^\mathfrak{r}_{m}b^\mathfrak{r}_m\!-\!h^\mathfrak{i}_{m}b^\mathfrak{i}_m)\!+\!{z}^\mathfrak{r}_{k,i} \!+\! j\sum_{m=1}^M(h^\mathfrak{r}_{m}b^\mathfrak{i}_m\!+\!h^\mathfrak{i}_{m}b^\mathfrak{r}_m)\!+\!j{z}^\mathfrak{i}_{k,i},
\end{eqnarray*}
where ${z}^\mathfrak{r}_{k,i}$, ${z}^\mathfrak{i}_{k,i}$ $\sim\mathcal{N}(0,\frac{N_0}{2d_k})$.
Thus, the likelihood function $f(y_{k,i}|\bm{w_{k,i}})$ is Gaussian, giving
\begin{eqnarray*}
f(y_{k,i}|&&\hspace{-0.65cm}\bm{w_{k,i}})\! \propto\! \exp\left\{\!\!-\frac{d_k}{N_0}\left[y^\mathfrak{r}_{k,i}\!-\!\!\sum_m (h^\mathfrak{r}_{m}b^\mathfrak{r}_m\!-\!h^\mathfrak{i}_{m}b^\mathfrak{i}_m)\right]^2 \!\right\} \\
&&\hspace{-0.6cm} \times \exp\left\{-\frac{d_k}{N_0}\left[y^\mathfrak{i}_{k,i}-\sum_m (h^\mathfrak{r}_{m}b^\mathfrak{i}_m+h^\mathfrak{i}_{m}b^\mathfrak{r}_m)\right]^2 \right\}.
\end{eqnarray*}

When we pass the information bottom up, $y_{k,i}$ is our observation (hence a constant) and $\bm{w_{k,i}}$ is the variable. Therefore, $f_b(\bm{w_{k,i}})=f(y_{k,i}|\bm{w_{k,i}})$.
After some manipulations, we can write $f_b(\bm{w_{k,i}})$ as a $2M$-dimensional Gaussian distribution:
\begin{eqnarray}\label{eq:fb}
f_b(\bm{w_{k,i}})\propto \mathcal{N}(\bm{w_{k,i}}, \bm{\eta_b}, \bm{\Sigma_b}),
\end{eqnarray}
where $\bm{\eta_b}$ and $\bm{\Sigma_b}$ are defined as
\begin{eqnarray}\label{eq:fb_params}
\bm{\eta_b} = \frac{2d_k}{N_0}
\begin{bmatrix}
\bm{\beta_1} \\
\bm{\beta_2}
\end{bmatrix}
\begin{bmatrix}
y^\mathfrak{r}_{k,i} \\
y^\mathfrak{i}_{k,i}
\end{bmatrix},~~
\bm{\Sigma_b} = \frac{2d_k}{N_0}
\begin{bmatrix}
\bm{\beta_1}\bm{\beta_1}^\top & \bm{\beta_1}\bm{\beta_2}^\top \\
\bm{\beta_2}\bm{\beta_1}^\top & \bm{\beta_1}\bm{\beta_1}^\top
\end{bmatrix},
\end{eqnarray}
and the matrices $\bm{\beta_1}$ and $\bm{\beta_2}$ are composed of channel coefficients as follows:
\begin{eqnarray*}
\bm{\beta_1}=
\begin{bmatrix}
h^\mathfrak{r}_{1} & h^\mathfrak{i}_{1} \\
h^\mathfrak{r}_{2} & h^\mathfrak{i}_{2} \\
\vdots & \vdots \\
h^\mathfrak{r}_{M} & h^\mathfrak{i}_{M} \\
\end{bmatrix},~~
\bm{\beta_2}=
\begin{bmatrix}
-h^\mathfrak{i}_{1} & h^\mathfrak{r}_{1} \\
-h^\mathfrak{i}_{2} & h^\mathfrak{r}_{2} \\
\vdots & \vdots \\
-h^\mathfrak{i}_{M} & h^\mathfrak{r}_{M} \\
\end{bmatrix}.
\end{eqnarray*}

In \eqref{eq:fb_params}, we have assumed that the dimensionality of $\bm{w_{k,i}}$ is $2M$, that is, $y_{k,i}=y_k[i]$ is related to $M$ complex variables. However, this is only valid when the number of neighbor symbols of $y_k[i]$ is $M$.
As shown in \eqref{eq:neighbors}, $\left|\mathcal{V}(y_k[i])\right|=M$ only when $1< i\leq L$. Thus, we can compute $\bm{\eta_b}$ and $\bm{\Sigma_b}$ by \eqref{eq:fb_params} only when $1< i\leq L$.

For the boundary samples ($i=1$, $L+1$) whose neighbor symbols are less than $M$, we further multiply the parameters $\bm{\eta_b}$ and $\bm{\Sigma_b}$ in \eqref{eq:fb_params} by an indicator vector $\bm{\gamma}$ and an indicator matrix $\bm{\Gamma}$, respectively, to ensure that $f_b(\bm{w_{k,i}})$ does not contain information about the symbols that do not belong to $\mathcal{V}(y_k[i])$. The general form of $\bm{\eta_b}$ and $\bm{\Sigma_b}$ are
\begin{eqnarray}\label{eq:fb_params2}
&& \bm{\eta_b} = \frac{2d_k}{N_0}
\begin{bmatrix}
\bm{\beta_1} \\
\bm{\beta_2}
\end{bmatrix}
\begin{bmatrix}
y^\mathfrak{r}_{k,i} \\
y^\mathfrak{i}_{k,i}
\end{bmatrix}\circ \bm{\gamma_k},  \\\label{eq:fb_params3}
&& \bm{\Sigma_b} = \frac{2d_k}{N_0}
\begin{bmatrix}
\bm{\beta_1}\bm{\beta_1}^\top & \bm{\beta_1}\bm{\beta_2}^\top \\
\bm{\beta_2}\bm{\beta_1}^\top & \bm{\beta_1}\bm{\beta_1}^\top
\end{bmatrix} \circ \bm{\Gamma_k},
\end{eqnarray}
where $\circ$ is an elementwise multiplication. The indicator vector $\bm{\gamma}$ and the indicator matrix $\bm{\Gamma}$ are defined as follows.

First, for the first $M$ samples (i.e., $i=1$), we have $\left|\mathcal{V}(y_k[i])\right|=k$ from \eqref{eq:neighbors}. Thus, we define
\begin{eqnarray*}
\bm{\gamma_k}=
\begin{bmatrix}
\begin{smallmatrix}
\bm{1}_{k\times 1} \\
\bm{0}_{(M\!-\!k)\times 1} \\
\bm{1}_{k\times 1} \\
\bm{0}_{(M\!-\!k)\times 1} \\
\end{smallmatrix}
\end{bmatrix},
\end{eqnarray*}
\begin{eqnarray*}
\bm{\Gamma_k}=
\begin{bmatrix}
\begin{smallmatrix}
\bm{1}_{k\times k}          &     \bm{0}_{k\times (M\!-\!k)}        &  \bm{1}_{k\times k}          &     \bm{0}_{k\times (M\!-\!k)}  \\
\bm{0}_{(M\!-\!k)\times k}   &      \bm{0}_{(M\!-\!k)\times (M\!-\!k)}   &      \bm{0}_{(M\!-\!k)\times k}   &      \bm{0}_{(M\!-\!k)\times (M\!-\!k)}   \\
\bm{1}_{k\times k}          &     \bm{0}_{k\times (M\!-\!k)}        &  \bm{1}_{k\times k}          &     \bm{0}_{k\times (M\!-\!k)}  \\
\bm{0}_{(M\!-\!k)\times k}   &      \bm{0}_{(M\!-\!k)\times (M\!-\!k)}   &      \bm{0}_{(M\!-\!k)\times k}   &      \bm{0}_{(M\!-\!k)\times (M\!-\!k)}
\end{smallmatrix}
\end{bmatrix},
\end{eqnarray*}
where $\bm{1}$ and $\bm{0}$ are all-ones and all-zero matrices with subscripts denoting their dimensions.

Second, for the last $M$ samples (i.e., $i=L+1$), we have $\left|\mathcal{V}(y_k[i])\right|=M-k$ from \eqref{eq:neighbors}. Thus, we define
\begin{eqnarray*}
\bm{\gamma_k}=
\begin{bmatrix}
\begin{smallmatrix}
\bm{0}_{k\times 1} \\
\bm{1}_{(M\!-\!k)\times 1} \\
\bm{0}_{k\times 1} \\
\bm{1}_{(M\!-\!k)\times 1} \\
\end{smallmatrix}
\end{bmatrix},
\end{eqnarray*}
\begin{eqnarray*}
\bm{\Gamma_k}=
\begin{bmatrix}
\begin{smallmatrix}
\bm{0}_{k\times k}          &     \bm{0}_{k\times (M\!-\!k)}        &  \bm{0}_{k\times k}          &     \bm{0}_{k\times (M\!-\!k)}  \\
\bm{0}_{(M\!-\!k)\times k}   &      \bm{1}_{(M\!-\!k)\times (M\!-\!k)}   &      \bm{0}_{(M\!-\!k)\times k}   &      \bm{1}_{(M\!-\!k)\times (M\!-\!k)}   \\
\bm{0}_{k\times k}          &     \bm{0}_{k\times (M\!-\!k)}        &  \bm{0}_{k\times k}          &     \bm{0}_{k\times (M\!-\!k)}  \\
\bm{0}_{(M\!-\!k)\times k}   &      \bm{1}_{(M\!-\!k)\times (M\!-\!k)}   &      \bm{0}_{(M\!-\!k)\times k}   &      \bm{1}_{(M\!-\!k)\times (M\!-\!k)}
\end{smallmatrix}
\end{bmatrix}.
\end{eqnarray*}

Finally, for all other samples ($1<i\leq L$), we simply set
\begin{eqnarray*}
\bm{\gamma_k}= \bm{1}_{2M\times 1},~~\bm{\Gamma_k}= \bm{1}_{2M\times 2M}.
\end{eqnarray*}
This is consistent with \eqref{eq:fb_params}.

2) Next, we consider the message $f_t(\bm{w_{k,i}})$ passed from the top. This message is the prior information of $\bm{w_{k,i}}$ and only added when $k=M$ (see Fig.~\ref{fig:5}). Therefore, we can write this message as
\begin{eqnarray*}
f_t(\bm{w_{k,i}})= \mathbbm{1}_{k=M}f_t(\bm{w_{M,i}}) + (1-\mathbbm{1}_{k=M})\bm{1}_{2M\times 1},
\end{eqnarray*}
that is, $f_t(\bm{w_{k,i}})$ is $f_t(\bm{w_{M,i}})$ when $k= M$ and an all-ones vector otherwise. In particular,
\begin{eqnarray}\label{eq:ft}
f_t(\bm{w_{M,i}})\propto \mathcal{N}(\bm{w_{M,i}}, \bm{\mu_t}, \bm{\Sigma_t}),
\end{eqnarray}
where
\begin{eqnarray*}
\hspace{-0.5cm}&& \bm{\mu_t}=\left[
\widehat{\mathbb{E}}_1^\mathfrak{r}, \widehat{\mathbb{E}}_2^\mathfrak{r}, ..., \widehat{\mathbb{E}}_M^\mathfrak{r},\widehat{\mathbb{E}}_1^\mathfrak{i}, \widehat{\mathbb{E}}_2^\mathfrak{i}, ..., \widehat{\mathbb{E}}_M^\mathfrak{i}\right]^\top,  \\
\hspace{-0.5cm}&& \bm{\Sigma_t} =
\frac{1}{2}\text{diag}\left(\widehat{\mathbb{D}}_1,\widehat{\mathbb{D}}_2,...,\widehat{\mathbb{D}}_M,\widehat{\mathbb{D}}_1,\widehat{\mathbb{D}}_2,...,\widehat{\mathbb{D}}_M\right).
\end{eqnarray*}
It is easy to transform the moment form of $f_t(\bm{w_{M,i}})$ to the canonical form by
\begin{eqnarray*}
f_t(\bm{w_{M,i}})\propto \mathcal{N}\left(\bm{w_{M,i}}; \bm{\eta_t} = \bm{\Sigma_t^\dagger \mu_t}, \bm{\Lambda_t=\Sigma_t^\dagger} \right).
\end{eqnarray*}

3) The third message, denoted by $f_\ell(\bm{w_{k,i}})$ in Fig.~\ref{fig:6}, is the message passed from $\bm{w_{k-1,i}}$ on the left. This message is obtained in the same way as $f_\ell(\bm{w_{k+1,i}})$ and we will analyze it at the end of forward message passing. For now, let us assume it is Gaussian and denote it by
\begin{eqnarray}\label{eq:fl}
f_\ell(\bm{w_{k,i}})\propto \mathcal{N}\left(\bm{w_{k,i}}; \bm{\eta_\ell}, \bm{\Lambda_\ell} \right).
\end{eqnarray}

4) As per the sum-product rule, the message out of a local function along an edge is the product of all incoming messages to this local function along all other edges. Thus, the message out of the equality function ``$=$'', denoted by $f_r(\bm{w_{k,i}})$ in Fig.~\ref{fig:6}, can be obtained by
\begin{eqnarray}
f_r(\bm{w_{k,i}})=f_b(\bm{w_{k,i}})f_t(\bm{w_{k,i}})f_\ell(\bm{w_{k,i}}).
\end{eqnarray}
This is the ``product'' step of the sum-product algorithm. From Lemma \ref{lemma2}, we know $f_r(\bm{w_{k,i}})$ is a Gaussian distribution, and
\begin{eqnarray}\label{eq:fr}
f_r(\bm{w_{k,i}})\propto \mathcal{N}\left(\bm{w_{k,i}}; \bm{\eta_r}, \bm{\Lambda_r} \right),
\end{eqnarray}
where $\bm{\eta_r}=\bm{\eta_b} +\bm{\eta_\ell}$, $\bm{\Lambda_r}=\bm{\Lambda_b}+ \bm{\Lambda_\ell}$. We emphasize that this message is an aggregation of all the known information about $\bm{w_{k,i}}$ from the left side of the graph.

The next step is to pass the message $f_r(\bm{w_{k,i}})$ through the compatibility function $\delta$. Notice that the compatibility function connects two different variables: on the LHS, the variable associated with the edge is $\bm{w_{k,i}}$; on the RHS, the variable associated with the edge is $\bm{w_{k+1,i}}$. Therefore, we have to integrate $f_r(\bm{w_{k,i}})$ over all variates that are in $\bm{w_{k,i}}$ but not in $\bm{w_{k+1,i}}$. This is the ``sum'' step of the sum-product algorithm.

Notice that the common variates of $\bm{w_{k,i}}$ and $\bm{w_{k+1,i}}$ are
\begin{eqnarray*}
\bm{w_\cap}=\left(b^\mathfrak{r}_1,...,b^\mathfrak{r}_k,b^\mathfrak{r}_{k+2},...,b^\mathfrak{r}_M,
b^\mathfrak{i}_1,...,b^\mathfrak{i}_k,b^\mathfrak{i}_{k+2},...,b^\mathfrak{i}_M \right),
\end{eqnarray*}
and the two different variates are $b^\mathfrak{r}_{k+1}$ and $b^\mathfrak{i}_{k+1}$ -- in $\bm{w_{k+1,i}}$, these two variates are $c^\mathfrak{r}_{k+1}$ and $c^\mathfrak{i}_{k+1}$.

Let us integrate $f_r(\bm{w_{k,i}})$ over $b^\mathfrak{r}_{k+1}$ and $b^\mathfrak{i}_{k+1}$, giving,
\begin{eqnarray*}
f(\bm{w_{\cap}})=\int_{b^\mathfrak{r}_{k+1}}\int_{b^\mathfrak{i}_{k+1}}f_r(\bm{w_{k,i}}) d b^\mathfrak{r}_{k+1} d b^\mathfrak{i}_{k+1}.
\end{eqnarray*}
As per Lemma \ref{lemma1}, $f(\bm{w_{\cap}})$ is also Gaussian. In particular, if we write $f_r(\bm{w_{k,i}})$ and $f(\bm{w_{\cap}})$ in moment form as
\begin{eqnarray*}
f_r(\bm{w_{k,i}}) \hspace{-0.2cm}& \propto &\hspace{-0.2cm} \mathcal{N}\left(\bm{w_{k,i}}; \bm{\mu_r}, \bm{\Sigma_r} \right), \\
f_r(\bm{w_{\cap}})\hspace{-0.2cm}& \propto &\hspace{-0.2cm} \mathcal{N}\left(\bm{w_{\cap}}; \bm{\mu_\cap}, \bm{\Sigma_\cap} \right),
\end{eqnarray*}
then $\bm{\mu_\cap}$ can be obtained by deleting the ($k+1$)-th and ($k+1+M$)-th rows of $\bm{\mu_r}$; $\bm{\Sigma_\cap}$ can be obtained by deleting the ($k+1$)-th and ($k+1+M$)-th rows and columns of $\bm{\Sigma_r}$.

However, $\bm{w_{\cap}}$ is a ($2M-2$) dimensional variable. To obtain $f_\ell(\bm{w_{k+1,i}})$, we have to expand the dimensionality of $\bm{w_{\cap}}$ by adding $c^\mathfrak{r}_{k+1}$ and $c^\mathfrak{i}_{k+1}$ in the ($k+1$)-th and ($k+1+M$)-th positions. After expansion, $f_\ell(\bm{w_{k+1,i}})$ is still multivariate Gaussian:
\begin{eqnarray}\label{eq:fl_right}
f_\ell(\bm{w_{k+1,i}})\propto \mathcal{N}\left(\bm{w_{k+1,i}}; \bm{\mu_\ell}, \bm{\Sigma_\ell} \right),
\end{eqnarray}
where $\bm{\mu_\ell}$ can be obtained by adding two zeros to $\bm{\mu_\cap}$; and $\bm{\Sigma_\ell}$ can be obtained by adding two all-zero rows and two all-zero columns to $\bm{\Sigma_\cap}$.

To summarize, we have shown that all the messages involved in the forward message passing are $2M$-dimensional multivariate Gaussians and can be parameterized by \eqref{eq:fb}, \eqref{eq:ft}, \eqref{eq:fl}, \eqref{eq:fr}, \eqref{eq:fl_right}, respectively.

{\it \textbf{Backward message passing}} -- Our tree structure is symmetric. Thus, backward message passing is symmetric to forward message passing. As shown in Fig.~\ref{fig:6}, to pass the messages from $\bm{w_{k+1,i}}$ and $\bm{w_{k,i}}$, we first compute three incoming messages $f_b(\bm{w_{k+1,i}})$, $f_t(\bm{w_{k+1,i}})$, and $f^\prime_r(\bm{w_{k+1,i}})$, where $f_b(\bm{w_{k+1,i}})$ and $f_t(\bm{w_{k+1,i}})$ are the same as that in the forward message passing and $f^\prime_r(\bm{w_{k+1,i}})$ is the message passed from $\bm{w_{k+2,i}}$ on the right.

Then, $f^\prime_\ell(\bm{w_{k+1,i}})$ and $f^\prime_r(\bm{w_{k,i}})$ are computed from ``product'' and ``sum'', respectively, by
\begin{eqnarray}
\label{eq:backward1}
f^\prime_\ell(\bm{w_{k+1,i}}) \hspace{-0.2cm} &=& \hspace{-0.2cm} f_b(\bm{w_{k+1,i}})f_t(\bm{w_{k+1,i}})f^\prime_r(\bm{w_{k+1,i}}), \\
\label{eq:backward2}
f^\prime_r(\bm{w_{k,i}})  \hspace{-0.2cm} &=&  \hspace{-0.2cm} \int_{c^\mathfrak{r}_{k+1}}\int_{c^\mathfrak{i}_{k+1}}f^\prime_\ell(\bm{w_{k,i}}) d c^\mathfrak{r}_{k+1} d c^\mathfrak{i}_{k+1}.
\end{eqnarray}

\begin{figure}[t]
  \centering
  \includegraphics[width=0.4\columnwidth]{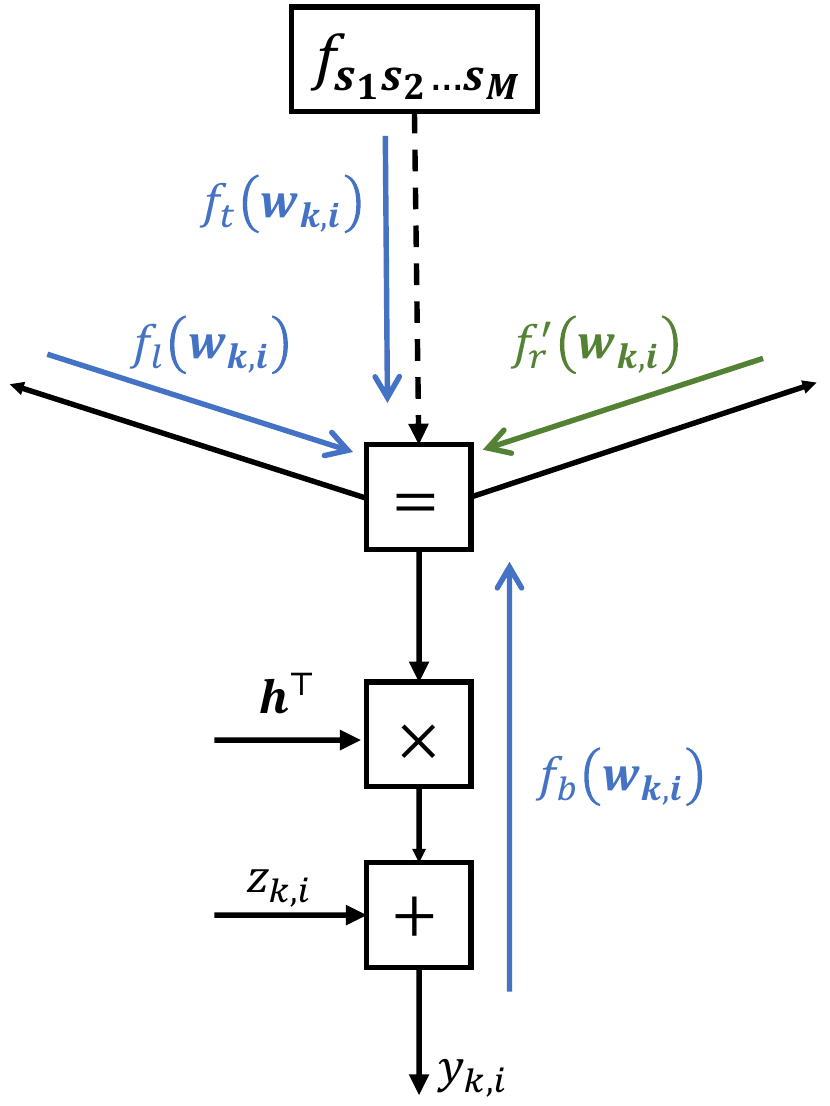}\\
  \caption{The marginalization process in the sum-product algorithm.}
\label{fig:7}
\end{figure}

{\it \textbf{Marginalization}} -- After one forward message passing from left to right and one backward message passing from right to left, the marginal posterior distribution of each variable $\bm{w_{k,i}}$ converges and can be computed by
\begin{equation}\label{eq:mar}
{\displaystyle
f(\bm{w_{k,i}}|\bm{y}) \!=\! f_b(\bm{w_{k,i}})f_t(\bm{w_{k,i}})f_\ell(\bm{w_{k,i}})f^\prime_r(\bm{w_{k,i}}),}
\end{equation}
as illustrated in Fig.~\ref{fig:7}. Therefore, $f(\bm{w_{k,i}}\allowbreak|\allowbreak\bm{y})$ is a $2M$-dimensional real Gaussian distribution. In particular, if we write the four messages on the RHS of \eqref{eq:mar} in the canonical form, then the canonical parameters of $f(\bm{w_{k,i}}|\bm{y})$ is the sum of them.

Recall that
$\bm{w_{k,i}}=\allowbreak(b^\mathfrak{r}_1,\allowbreak ...,\allowbreak b^\mathfrak{r}_M,\allowbreak  b^\mathfrak{i}_1,...,\allowbreak b^\mathfrak{i}_M)$
is a $2M$ dimensional real random variable, where $b^\mathfrak{r}_m$ and $b^\mathfrak{i}_m$ are the real and imaginary parts of the $m$-th complex element of
$\bm{W_{k,i}}=\mathcal{V}(y_k[i])=(\allowbreak s_1[i],\allowbreak s_2[i],\allowbreak ...,\allowbreak s_k[i],\allowbreak s_{k+1}[i-1],\allowbreak s_{k+2}[i-1],...,\allowbreak s_M[i-1])$, thus, $f(\bm{W_{k,i}}|\bm{y})$ is an $M$-dimensional complex Gaussian distribution.

Let $k = M$, we have $\bm{W_{M,i}}=(s_1[i],s_2[i],...,s_M[i])=\bm{s[i]}$. This means
\begin{eqnarray}\label{eq:marginal}
f(\bm{s[i]}|\bm{y})=f(\bm{W_{k,i}}|\bm{y})
\end{eqnarray}
is an $M$-dimensional complex Gaussian, the mean and covariance of which can be computed from \eqref{eq:mar}.

\bibliographystyle{IEEEtran}
\bibliography{References}

%
%
\end{document}